\documentclass[12pt,preprint]{aastex}

\usepackage{eso-pic}
\usepackage{graphicx}
\usepackage{color}
\usepackage{type1cm}
\makeatletter
  \AddToShipoutPicture{%
    \setlength{\@tempdimb}{.5\paperwidth}%
    \setlength{\@tempdimc}{.5\paperheight}%
    \setlength{\unitlength}{1pt}%
    \put(\strip@pt\@tempdimb,\strip@pt\@tempdimc){%
    }
}
\makeatother

\newcommand{\um}{\ensuremath{\rm {\mu}m}}
\newcommand{\kepler}{{\it Kepler}}

\slugcomment{Submitted for publication in the Astrophysical Journal}
\shortauthors{McCullough}

\received{2006 February XX}
\begin{document}

\title{Models of Polarized Light from Oceans and Atmospheres of Earth-like Extrasolar Planets}

\author{P. R. McCullough\altaffilmark{1}}

\email{pmcc@stsci.edu}

\altaffiltext{1}{Space Telescope Science Institute, 3700 San Martin Dr., Baltimore MD 21218}

\begin{abstract}
Specularly reflected light, or glint, from an ocean surface may
provide a useful observational tool for studying extrasolar terrestrial planets.
Detection of sea-surface glints would differentiate 
ocean-bearing terrestrial planets, i.e. those similar
to Earth, from other terrestrial extrasolar planets.
Sea-surface glints are localized and linearly polarized. The brightness
and degree of polarization of both sea-surface glints and atmospheric Rayleigh
scattering are strong functions of the phase angle of the extrasolar
planet.
The difference of the two orthogonal linearly polarized reflectances may
be an important observational signature of Rayleigh scattering or glint.
The difference attributable to Rayleigh scattering peaks near quadrature,
i.e. near maximum elongation of a circular orbit.
The difference attributable to glint peaks at crescent phase, and in crescent
phase the total (unpolarized) reflectance of the glint is also near maximum.
We present analytic and physical models of sea-surface glints.
We modify analytic expressions for the bi-directional reflectances
previously validated by satellite imagery of the Earth to account
for the fractional linear polarization
of sea-surface reflections and of Rayleigh scattering in the atmosphere.
We compare our models with Earth's total visual light and degree of
linear polarization as observed in the ashen light of the Moon, or Earthshine.
We predict the spatially-integrated reflected
light and its degree of polarization as functions of the diurnal cycle
and orbital phase of Earth and Earth-like planets of various imagined types.
The difference in polarized reflectances of Earth-like planets may increase
greatly 
the detectability of such planets in the glare of their host star.
Finally, sea-surface glints potentially may provide a practical means to map the
boundaries between oceans and continents on extrasolar planets.
\end{abstract}

\section{Introduction}

Terrestrial extrasolar planets have captured the human
imagination for much of history. 
Missions such as \kepler\ will search for planets in the
so-called ``habitable zone,'' in which the planet may support water in its
liquid state (Borucki et al. 2006).
The proximity of accomplishing
the goal of detecting terrestrial planets, especially in relatively large
numbers, begs the question, ``What next?''
One purpose of this paper is to inform the answer to that question.

We consider the next\footnote{Next in two senses of the word: next in time and next in significance.} important goal could be to identify
extrasolar planets that not only {\it could} support liquid water based
upon their size, mass, and proximity to their star, but also 
indeed demonstrably {\it do} have liquid on their surfaces, i.e. oceans,
and {\it do} have atmospheres of non-negligible column density.
Both oceans and atmospheres polarize light reflected from them. In this
paper we examine linear polarization as a potentially useful signature of
oceans and atmospheres of Earth-like extrasolar planets.

Some of the concepts in this paper have been described
independently by others.
Hough \& Lucas (2003) and Stam, Hovenier, \& Waters (2004)
noted the efficacy of polarization to reduce the
glare of a nearly unpolarized star compared to the Rayleigh-scattered
polarized light of a hot Jupiter. 
Seager et al. (2000), Saar \& Seager (2003), and Stam \& Hovenier (2005)
examined the Rayleigh-scattered light of a hot Jupiter in greater detail. 
Williams \& Gaidos (2004) and Gaidos et al. (2006) examined the
unpolarized variability of
the sea-surface glint from an Earth-like extrasolar planet, and Williams
(2006) has proposed that a satellite of a
extrasolar gas giant planet could be brighter than the planet itself due to
the much lower albedo of the planet than that of the satellite.
Stam \& Hovenier (2006) independently examined the observability
of the polarized signatures of an Earth-like extrasolar planet, including
Rayleigh scattering and sea-surface glint.
Schneider (2006) had similar ideas.

Kuchner (2003) and L{\'e}ger et al. (2004) have proposed that ``ocean planets''
may form from ice planets that migrate inward and melt;
the surfaces of these planets
would be liquid water exclusively, i.e. no continents. Although Kuchner
expects the liquid surface to be entirely obscured by a thick steam atmosphere,
L{\'e}ger et al. model a liquid ocean under either an obscuring cloud or
a clear atmosphere.
In the former case,
the specular reflection from the ocean would not be visible, and the
evidence of liquid water would be indirect: the mean density could be
inferred from transits and radial velocities, and the surface temperature
could be inferred
from the star's incident flux and the planet's albedo or from the planet's
infrared emissivity.

In this paper, we explore the potential observational
signature of a liquid, especially water, on a planet's surface: the polarized
glint.
The observational signatures of
glints may be detectable in the orbital phase-dependent
and time-dependent spatially-integrated
reflected light of the extrasolar planet.
The phase dependency of the polarization
may allow future astronomers to infer a
specularly reflecting surface, and (less reliably) to measure the index of
refraction of the medium, n, and if n $\sim\ 1.3$ then
that might give some credence to speculation that the medium is liquid water.
However, remote sensing of distant worlds generally will
admit many possible interpretations, as the following example of
Titan illustrates.

West et al. (2005) interpreted the apparent lack of
near-infrared specular reflection from Titan, combined with the
specular signatures from Earth-based radar as evidence for no liquid
oceans on its surface at the locations they studied, circling Titan at
${\sim} -30$\arcdeg\ latitude. Liquid lakes have been detected on Titan
by observations of the Cassini orbiter, at latitudes
north of $+ 70$\arcdeg\ (Mitchell et al. 2006).
In general, a lack of a glint cannot prove that
liquids do not exist {\it anywhere} on an extrasolar planet's surface.
The example of Titan also shows that a glint is not necessarily indicative
of liquid {\it water}.

Neither radar nor highly spatially-resolved planetary
images, such as those of Titan, will be available for extrasolar planets.
However extrasolar planets could exhibit a wide range of phase angles, an
impossibility for Earth-based observations of the outer planets.
For example, using the phase dependency of polarization, Horak (1950)
showed the fractional polarization
of Venus' atmosphere is not large and is not due to Rayleigh scattering.

In Section \ref{sec:orrery} we analyze physical models of specularly
reflecting spheres. In Section \ref{sec:brdfs} we modify bi-directional
reflectances to account for linear polarization from two mechanisms:
Rayleigh-scattering and specular reflection. In Section \ref{sec:models}
we model Earth-like planets and demonstrate that the rotation of a planet
with surface features such
as continents and oceans should modulate the polarized reflectances in a simple
and predictable manner.
We also simulate the modulation of the polarized reflectances
due to orbital phase.
Section \ref{sec:earthshine} compares our models with observations of
Earthshine.
Section \ref{sec:discussion} discusses the technique of using the glint
as a tool for mapping extrasolar planets and estimates the glint's
observability with near-future instrumentation.

\section{Observations and Analysis of a Simple Orrery\label{sec:orrery}}

We constructed simple physical models of planets with uniform surfaces and
no atmospheres. The models are 2.5-inch diameter wooden spheres covered
uniformly with acrylic paint.  Sphere A is painted flat white; sphere B,
flat white with a shiny transparent acrylic overcoat; and sphere C, flat
black with an identical overcoat.  The acrylic overcoats of spheres B
and C specularly reflect the incident light. Sphere C is meant to model
a deep ocean, because it has a very low albedo for light that is not
specularly reflected from the first air-acrylic interface.  Sphere B
could represent the surface of a planet with a very shallow, sandy
ocean, or a mix of ice and water.  Each sphere is suspended in a dark
room by a wooden dowel, 23 inches (58 cm) from a incandescent filament
covered by a
1/4-inch (6-mm) diameter, white, translucent plastic dowel that makes the lamp's
radiance isotropic azimuthally, i.e. in the plane defined by the lamp, the
sphere, and the camera. That plane is the plane of incidence for specularly
reflected rays, and by convention,
the electric field of the s-component is perpendicular ($\perp$)
to that plane, and the electric field of the
p-component is parallel ($\parallel$) to the plane of incidence.
Each sphere was photographed with a quantitative
digital camera at angles corresponding to ``orbital'' phase of 30\arcdeg,
50\arcdeg, 70\arcdeg, 90\arcdeg, 110\arcdeg, 130\arcdeg, and 150\arcdeg.
At each phase angle, four pairs of exposures were taken of each sphere
at four angles (0\arcdeg, 45\arcdeg, 90\arcdeg, and 135\arcdeg) of
a linear polarizer in front of the digital camera's lens. 
At 0\arcdeg\ orientation, the polarizer transmitted the p-component
and at 90\arcdeg\ it transmitted the s-component.

Even in an otherwise darkened
room the light from the lamp illuminates the walls, floor,
and ceiling of the room which then illuminate the model ``planet.'' The
latter illumination is undesirable because no such illumination occurs
for a real planet. To compensate, we took matched exposure pairs: one
with and one without an occulting sphere placed between the ``planet''
and its ``star.''  By subtracting the image of the planet in ``eclipse''
from that of it illuminated directly, we effectively removed the unwanted,
indirect illumination.

The spheres can be modeled 
as a Lambertian surface of albedo A with or without
a specular reflection component superposed.\footnote{Because the specularly
reflected light is not available to be reflected from the
Lambertian surface underneath, superposition is only an approximation.}
The phase angle $\gamma$ is defined to be the
angle separating the illuminator and the detector measured from the center
of the sphere.
The phase function of the
integrated light for a Lambertian sphere is (Russell 1916),
\begin{equation}
\label{eq:lambert}
I(\gamma) = {A \over \pi} ( sin \gamma + (\pi - \gamma) cos \gamma ),
\end{equation}
which is normalized such that $I(\gamma = 0) = A$, i.e. the results 
are normalized to full phase and are proportional to the albedo A. 
With a consistent normalization, the phase function for a sphere with a 
perfect specularly reflecting surface is (Tousey 1957)
\begin{equation}
\label{eq:tousey}
I(\gamma) = {1 \over 4} \times {3 \over 2} = {3 \over 8}.
\end{equation}
Thus, the phase function for the specular reflectance from a sphere is
\begin{equation}
\label{eq:specular}
I(\gamma) = {3 \over 8} r,
\end{equation}
where r is a linear combination of the Fresnel reflection coefficients,
\begin{equation}
\label{eq:fresnel1}
r_s = {{ sin^2( \theta_i - \theta_t ) }\over { sin^2( \theta_i + \theta_t ) }}
\end{equation}
and
\begin{equation}
\label{eq:fresnel2}
r_p = {{ tan^2( \theta_i - \theta_t ) }\over { tan^2( \theta_i + \theta_t ) }},
\end{equation}
where the angle of incidence $\theta_i = \gamma / 2$, and
according to Snell's law, $n sin \theta_t  = sin \theta_i$, in which
n is the refractive index of the transmissive medium and $\theta_t$ is the
angle of the transmitted ray.\footnote{We approximate the refractive index of
air as that of vacuum.}
The appropriate linear combination depends on the
orientation of the linear polarizer in front of the lens of the digital camera.

Sphere A's ``flat'' painted surface is a good approximation to
a Lambertian surface. Sphere B's phase function is approximately that of
a Lambertian sphere with a specular-reflection component superposed.
The phase functions for spheres A and B are not shown because each is
nearly as expected analytically from
Equations \ref{eq:lambert}, \ref{eq:specular}, 
\ref{eq:fresnel1}, and \ref{eq:fresnel2}.

\begin{figure}
\plottwo{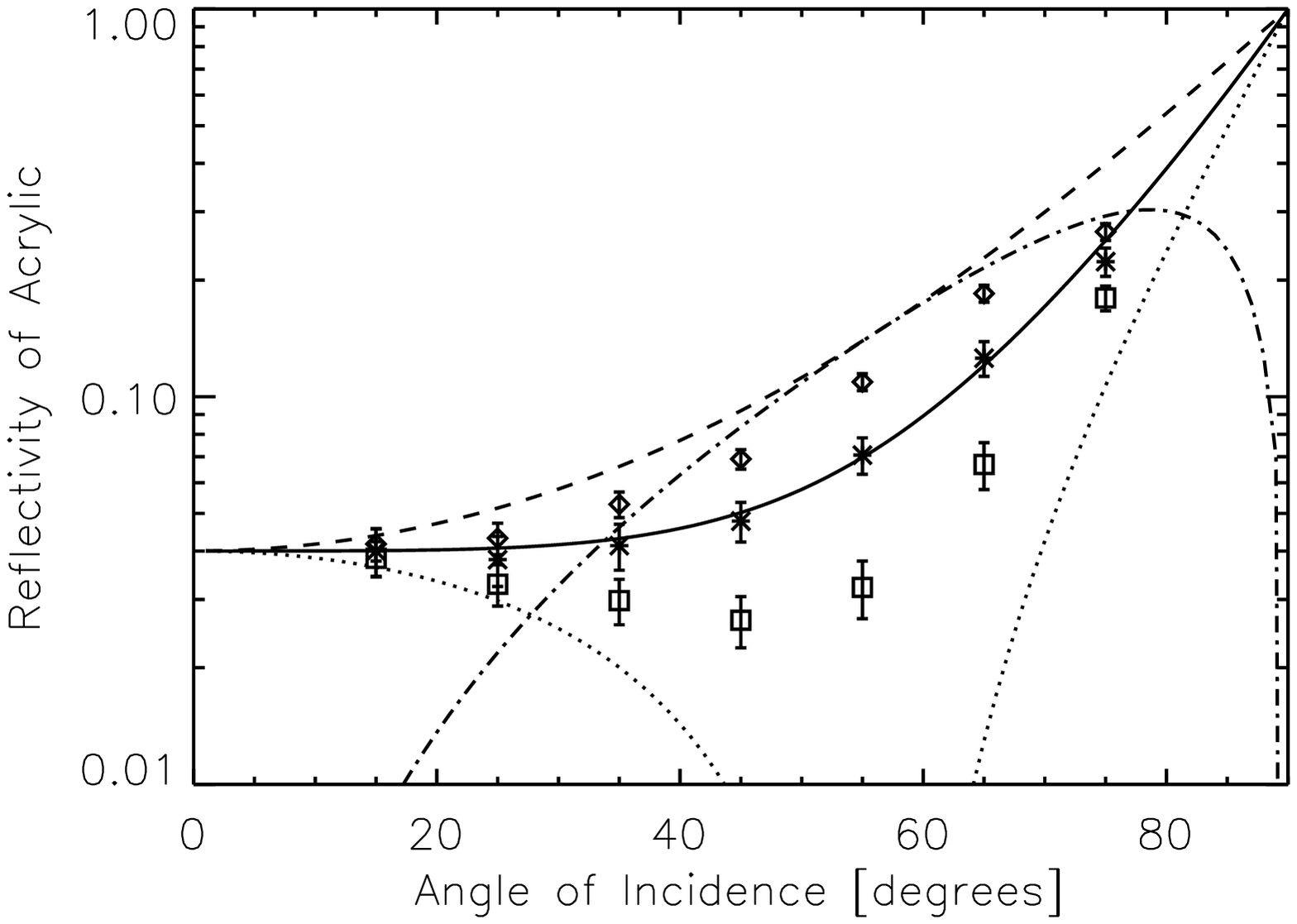}{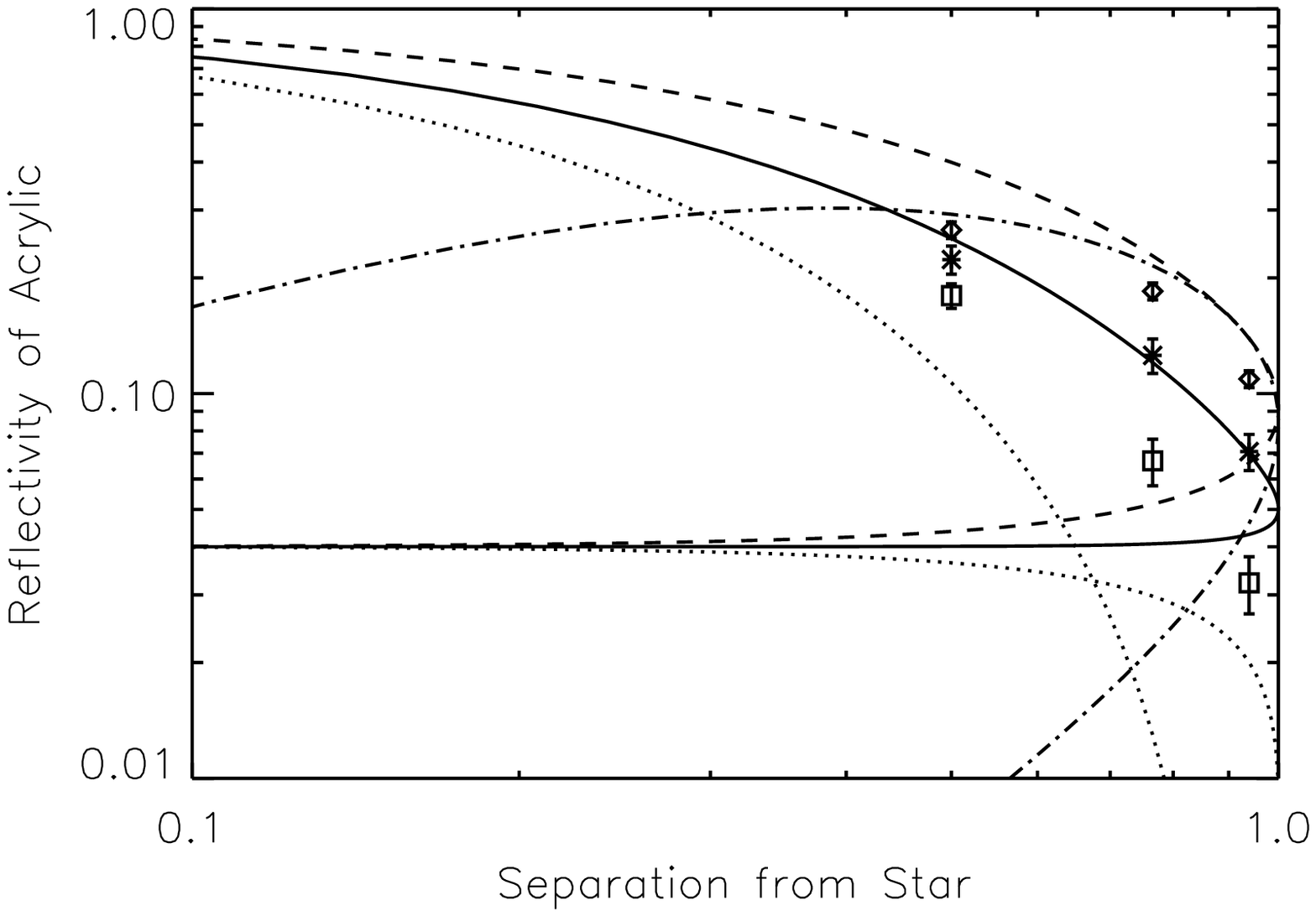}
\caption{Specular reflectivity of acrylic. Integrated reflectances
from a sphere painted flat black with a
shiny acrylic overcoat are plotted at each angle
in the two orthogonal s- and p-components of linear polarization, upper
and lower points respectively.
Each middle data point is the mean of the other two.  
We selected a single scale factor to match the observed mean reflectances
to the theoretical curve.
The dashed and dotted lines are the s- and p-components respectively,
for refractive index n = 1.50 from Equations
\ref{eq:fresnel1} (dashed) and \ref{eq:fresnel2} (dotted),
their mean (solid), and their difference (dashed-dotted).
The same curves are plotted with respect to angle of incidence
(left) and projected separation from the ``star,'' i.e.  the source of
illumination (right). In the latter, the separation is normalized to
the maximum separation, a circular orbit is assumed, and only points
with phase angle $\gamma \ge 110$\arcdeg, i.e. crescent phase,
are plotted, for clarity.
\label{fig:orrery1}}
\end{figure}

Because one purpose of this paper is to examine the specular reflectance
from an ocean that absorbs all the light transmitted below its surface,
we compare the observed and theoretical phase functions of sphere C,
which is painted flat black with a shiny transparent acrylic overcoat
(Figure \ref{fig:orrery1} and Section \ref{sec:uniform}). The
glint contributes significantly to the spatially integrated light from
sphere C at all 
phase angles and creates the large ratio of one linearly polarized
reflectances to the
other at phases corresponding to the Brewster angle of incidence.
Because the indices of refraction
for acrylic and water are 1.50 and 1.33 respectively,
the phase angles $\gamma_B$ of the theoretical nulls of the p-component are
112.6\arcdeg\ and 106.1\arcdeg, which are double the respective
Brewster angles, 56.3\arcdeg\ and 53.1\arcdeg, respectively.
For our painted spheres, the 
observed phase functions are approximately as predicted by Eq. \ref{eq:specular},
although the p-component of the specular reflectance is larger than
predicted, and the s-component is smaller than predicted.

\begin{figure}
\plottwo{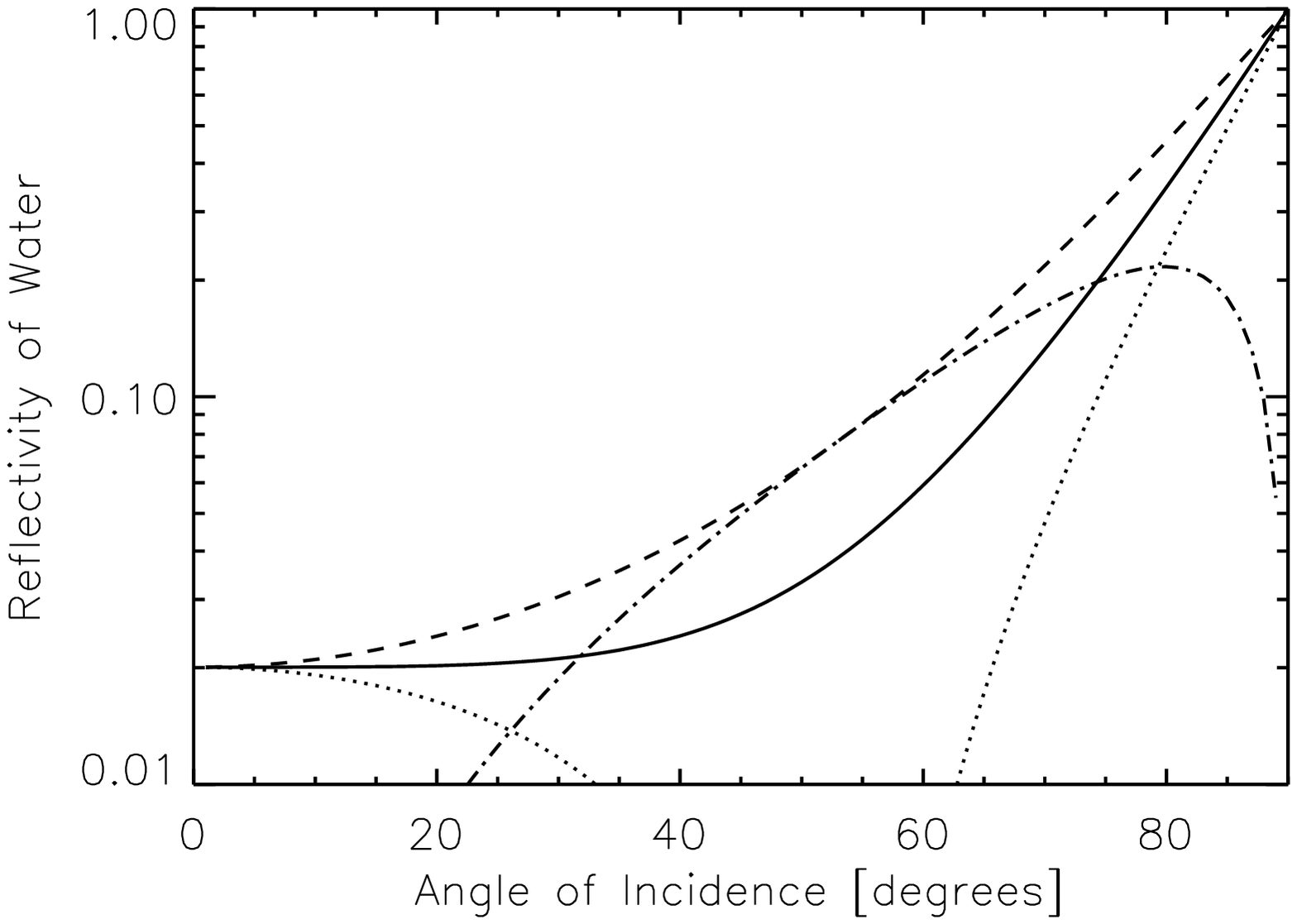}{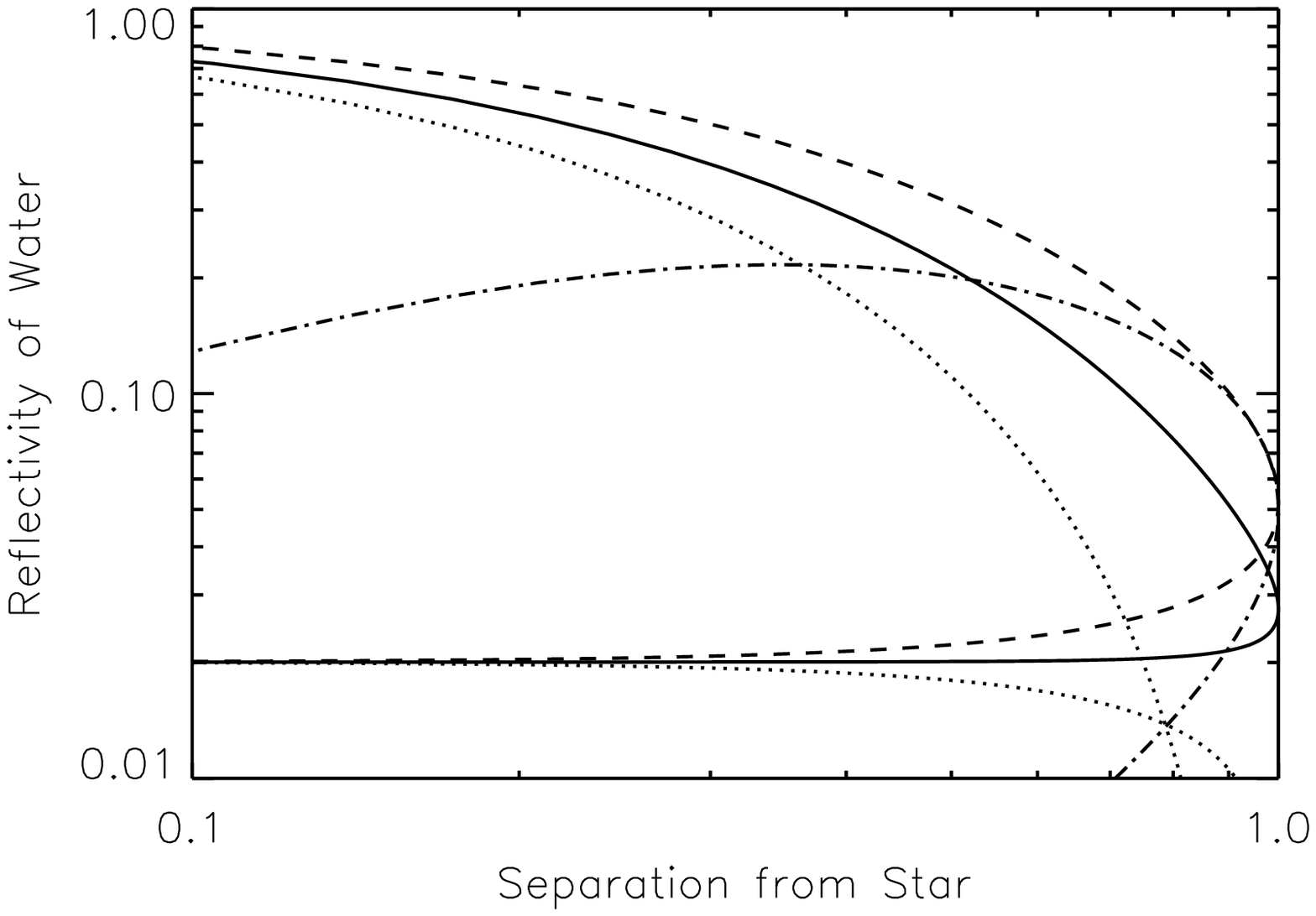}
\caption{Specular reflectivity of water. Same as Figure
\ref{fig:orrery1} except for an index of refraction n = 1.33. No
experimental data are plotted.
\label{fig:orrery2}}
\end{figure}

\begin{figure}
\plottwo{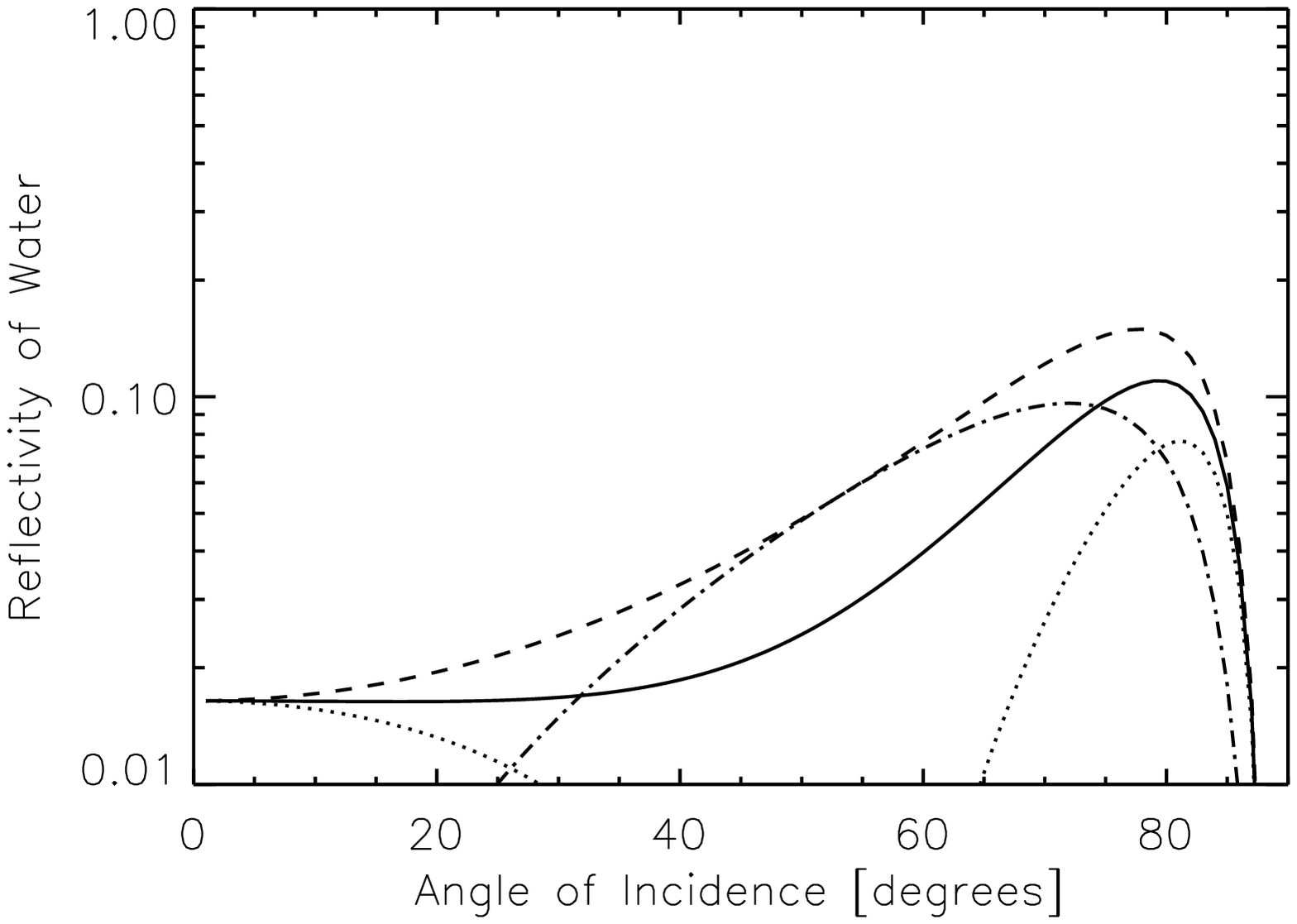}{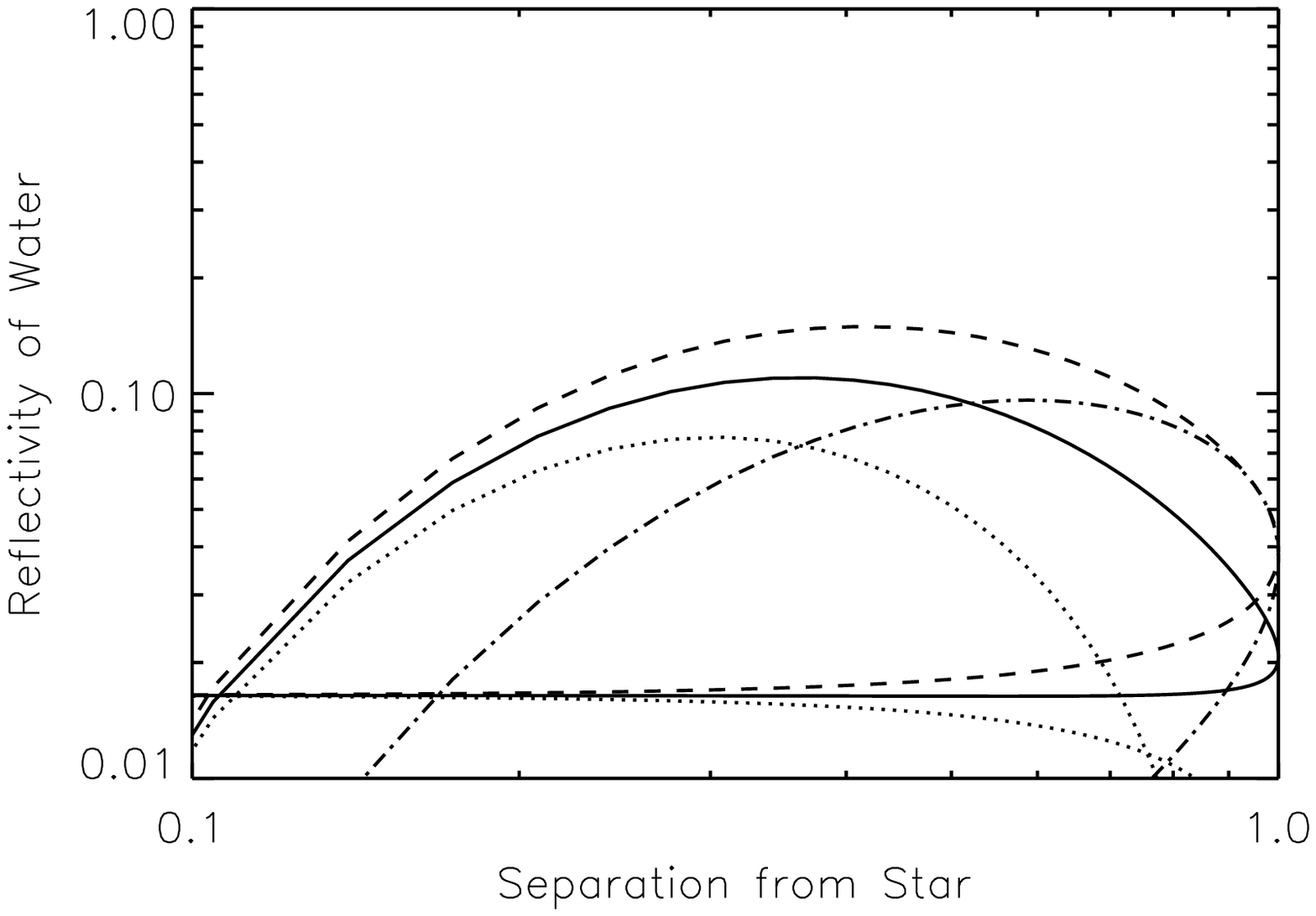}
\caption{Specular reflectivity of water, including atmospheric absorption.
Same as Figure
\ref{fig:orrery2} except reflectivities are attenuated by
exp(-0.1($sec \zeta_s + sec \zeta_o$)), where $sec \zeta_s$ and $sec \zeta_o$
are the airmasses of the star and the
observer, at their respective zenith angles, $\zeta_s$ and $\zeta_o$, at
the point of specular reflection.
\label{fig:orrery3}}
\end{figure}

\begin{figure}
\plotone{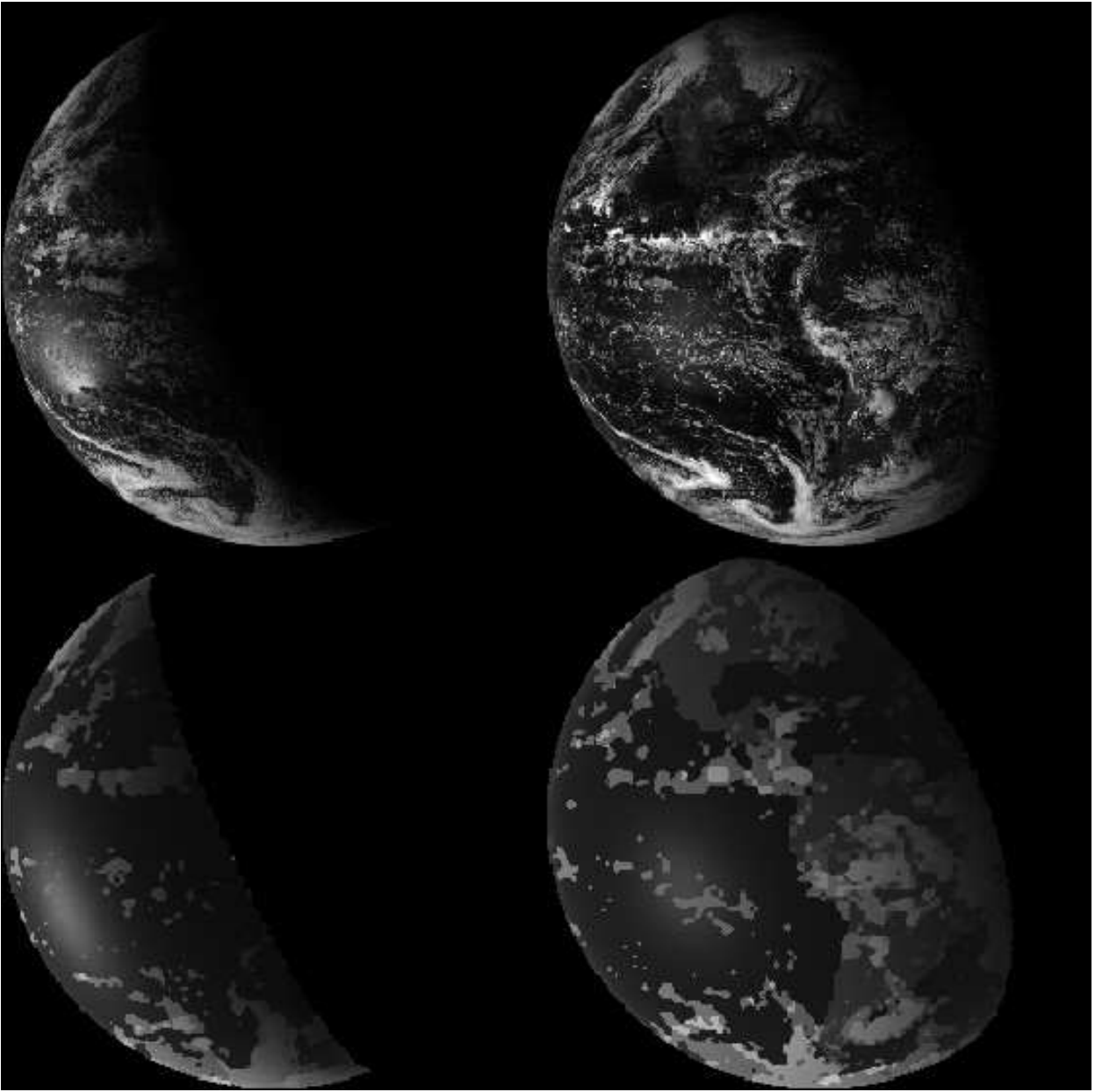}
\caption{Images from the geostationary satellite GOES EAST (above) taken
in unpolarized light on Dec 28, 2005 at 23h45m UTC (left) and 20h45m (right)
and corresponding numerical models for the same (below). 
The phase angle $\gamma = 100$\arcdeg\ (left) and 59\arcdeg\ (right).
The sea-surface glint is the bright (white) patch to the lower-left of center of
the Earth.
The Americas are visible in the image and model at right.
\label{fig:goesnone}}
\end{figure}

\begin{figure}
\plotone{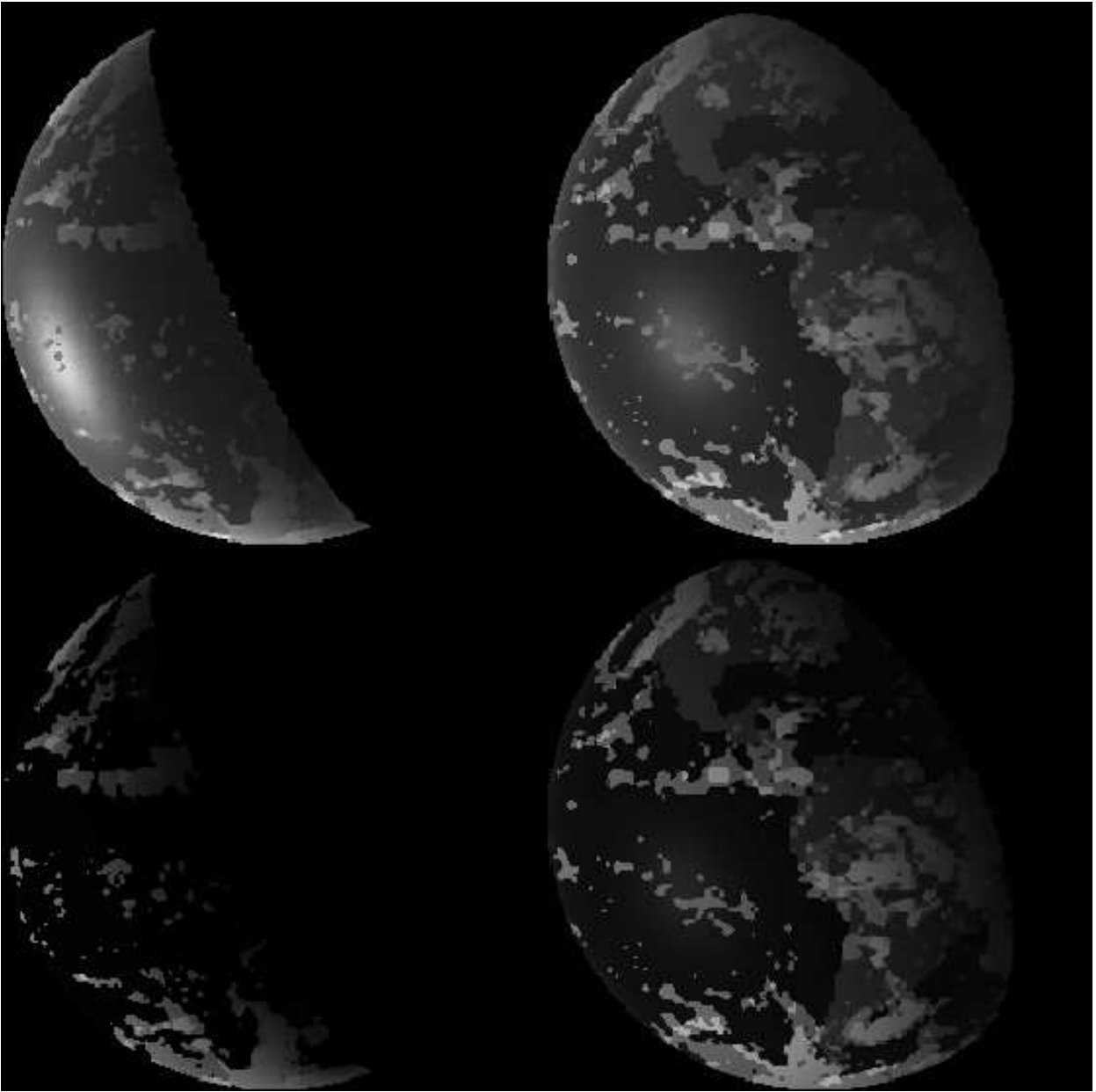}
\caption{Models images of the scenes from Figure \ref{fig:goesnone}
are shown for linear polarized components of light
($\perp$\ or s-component above; $\parallel$\ or p-component below).
\label{fig:goespolarized}}
\end{figure}

A simple physical model such as sphere C is useful to understand the
optical characteristics of specular reflection from a sphere 
(Figures \ref{fig:orrery1}-\ref{fig:goespolarized}):
1) the mean reflectance increases with angle of incidence, 2) the Brewster angle
occurs near maximum elongation of a circular orbit, 3) the {\it difference}
between the polarized reflectances is large for nearly all the ``crescent''
phases, $\gamma \ga 90$\arcdeg, and 4)
the glint elongates spatially at large angles of incidence.
We constructed the physical models in part to validate our numerical models. 
Similar models could provide
opportunities for teaching and outreach at many levels of sophistication,
from elementary school to graduate school.

The painted spheres do not model the atmospheric extinction. To account for
extinction in a cloudless, Earth-like atmosphere, we 
multiply the reflectance of Equation \ref{eq:specular}
(and Figure \ref{fig:orrery2}) by an optical opacity,
\begin{equation}
\label{eq:speculartimesextinction}
I^\prime(\gamma) = I(\gamma) e^{-0.1(sec\ \zeta_s + sec\ \zeta_o)},
\end{equation}
where $\zeta_s$ and $\zeta_o$
are the zenith angles of the star and the
observer, respectively, at the point of specular reflection toward the observer
(Figure \ref{fig:orrery3}). 
At the point of specular reflection, 
the star's zenith angle $\zeta_s$ equals
the angle of incidence $\theta_i$, which equals the angle of reflection, so
the airmasses are equal, $sec \zeta_s = sec \zeta_o$.
The extinction coefficient, 0.1, is appropriate for broadband visible light
with a solar spectrum (Hayes \& Latham 1975). Much of the extinction in a
clear atmosphere is due to scattering, not absorption. For astronomical
photometry of a star, whether light is lost to absorption or scattering is
immaterial. In the case of glint from an ocean, by reciprocity, a similar
circumstance applies to the upwelling light (i.e.
after reflection from the surface), traveling from the planet's surface to
the observer: scattered light is lost as effectively as absorbed light.
The same is true of downwelling light that is scattered upward. 
However, downwelling light that is forward scattered (downward) may still
contribute to the glint, because it may specularly reflect to the observer
from a different facet of another wave on the ocean's surface. 
In our approximation, we have
neglected that effect: scattered light is not permitted to reflect.
We have neglected also the interaction of the
polarization states of the scattered and specularly reflected light.

\section{Analytic Approximations for Polarized Bi-directional Reflectances\label{sec:brdfs}}

We use the analytic bi-directional reflectances of
Manalo-Smith et al. (1998, hereafter Paper 1) for visible light,
as calibrated from various scene types observed by satellites.
We modified the  bi-directional reflectances to account
for the fractional polarization of Rayleigh scattering in the clear
atmosphere and of sea-surface specular reflection.
The fractional polarization $p$ is defined from the intensities of the two
orthogonal, linearly polarized components,
\begin{equation}
\label{eq:pdef}
p = {{ I_s - I_p }\over { I_s + I_p }}.
\end{equation}
\subsection{Rayleigh Scattering}

For Rayleigh scattering the two components are
\begin{equation}
\label{eq:ray1}
I_s = k \times 1
\end{equation}
and
\begin{equation}
\label{eq:ray2}
I_p = k \times cos^2 \gamma
\end{equation}
where k is independent of $\gamma$ (Lang 1999).
Hence, the fractional polarization of Rayleigh scattering is
\begin{equation}
\label{eq:pray}
p_{Ray} = {{ sin^2 \gamma }\over { 1 + cos^2 \gamma }}.
\end{equation}
The polarized bi-directional reflectance attributable to Rayleigh scattering is
\begin{equation}
\label{eq:rray}
r_{Ray} = C_2 {{ 1 + cos^2 \gamma }\over { ( u u_0 )^{C_3} }} \ {{1 \pm p_{Ray}}\over{2}}
\end{equation}
where $C_2$ and $C_3$ are coefficients determined by analysis of
Earth Radiation Budget Experiment (ERBE) satellite data and $u$ and $u_0$ are direction cosines
(see Paper 1). The last term in Equation \ref{eq:rray},
$(1 \pm p_{Ray})/2$ partitions the unpolarized reflectance from
Paper 1 into its two polarized reflectances.

For low optical thickness, low surface albedo, and moderate
single-scattering optical depth, the degree of polarization $p_{Ray}$
of a Rayleigh
scattered atmosphere is nearly unity at phase angle $\gamma =90$\arcdeg,
and tapers to zero at phase angles 0\arcdeg\ and 180\arcdeg.
Kattawar \& Adams (1971) and Viik (1990)
have generalized that result for various optical thicknesses and
surface albedos.
Because we use a single-scattering model for analytic simplicity, our models
may over-estimate the fractional polarization due to Rayleigh scattering.
Multiple scattering would be more physically realistic and tend to
depolarize the emergent light by a few per cent
(e.g. Stam \& Hovenier 2005).
However, the fractional polarization
predicted by the approximations in this paper should not be considered
as upper limits,
simply because less (or more) cloud cover on an extrasolar
planet will increase (or decrease) its net polarization.

In the single-scattering Rayleigh model,
the fractional polarization $p_{Ray}$ is
independent of wavelength, for wavelengths much larger than the size of
the scattering particles. However, as Wolstencroft \& Breon (2005) have noted,
the fractional polarization of the spatially-integrated light of an Earth-like
planet should decrease with increasing optical wavelength due to the
strong wavelength dependence ($\lambda^{-4}$) of the Rayleigh-scattering cross
section (Jackson 1975) and the flat spectrum of clouds. 
Dollfus (1957) observed that the polarization of the Earthshine decreased
with increasing wavelength: p = 8.4\%, 5.4\%,  and 3.5\% at
$\lambda$ = 0.49 \um,  0.55 \um, and 0.63 \um,
respectively.\footnote{Values are for Earth at phase $\gamma = 70$\arcdeg,
i.e. a crescent moon.
Multiply p by $\sim$4 to obtain the fractional polarization of the light
of Earth without reflection from the Moon.}
Wolstencroft \& Breon (2005) predict fractional polarization of the Earth's
light at $\lambda = 0.443$ \um\ and at phase angle $\gamma = 90$\arcdeg\ to be
20\% to 28\%, or 0.5 to 0.7 times Dollfus' estimate or
our own (both 40\%; Section \ref{sec:earthshine}).

The radiances of the clear daytime sky and of Earthshine (Woolf et al. 2002)
both decrease with increasing wavelength. In each case, the cause is
attributed to atmospheric Rayleigh scattering, which is diluted by the
flat spectrum of clouds, causing the Earth to appear, in the words
of Sagan (1994), as a ``Pale Blue Dot.''

The radiance of the Rayleigh scattering in our model should be proportional to
$C_2 (\lambda/\lambda_{eff})^{-4}$, where $\lambda_{eff}$ is a
wavelength at which the coefficient $C_2$ is appropriate.
The coefficient $C_2$ is derived from observations from ERBE scanning
radiometers that have sensitivity over 0.2-3.8 $\mu$m, decreasing
gradually with wavelength for $\lambda < 1 \mu$m (Smith et al. 1986).
The ERBE radiometer's sensitivity to a $\lambda^{-4}$
Rayleigh-scattered blackbody approximation to the solar spectrum peaks
at $0.27 \mu$m, and
the weighted-mean wavelength equals $0.35\mu$m.
Modeling the telluric absorption by ozone as transparent ($\lambda > 0.35$\um)
or opaque ($\lambda < 0.35$\um) (see e.g. Hayes \& Latham 1975)
shifts the weighted mean-wavelength from $0.35\mu$m to $0.47\mu$m.
The equivalent weighted-mean wavelength for the human eye's scotopic
response (Wyszecki \& Stiles 1982) to Rayleigh-scattered sunlight is $0.49\mu$m.
Evidently, the observations of the clear atmosphere by the ERBE
radiometers and those by the human eye will be similar, and in Section
\ref{sec:earthshine} we compare such observations.

\subsection{Specular Reflection}

We partition the unpolarized reflectance for specular reflection
(again from Paper 1),
\begin{equation}
\label{eq:rsp}
r_{sp} = C_4 {{ C_5 - 1 }\over { ( u u_0 )^{1.5} (C_5 - cos \alpha )^2 }} \ {{1 \pm p_{sp}}\over{2}},
\end{equation}
into its two polarized reflectances with the last term, 
where the polarization of specular reflections from water $p_{sp}$ is
determined from 
Equations \ref{eq:specular}, \ref{eq:fresnel1}, \ref{eq:fresnel2},
and \ref{eq:pdef}. Like Equation \ref{eq:rray}, the other variables of
Equation \ref{eq:rsp} are discussed in Paper 1, so we
do not repeat that discussion for the sake of brevity.
In this case,
the specular reflection comes from an area on the sphere determined by the
distribution of tilts of the air-water interface (waves)
(Zeisse 1995; Takashima 1985).
If a facet of a wave is to specularly reflect the
star's light to the observer, then the angle of incidence of
the specular reflection on that facet should be half the
phase angle $\gamma$, as before for the idealized sphere. 

\section{Numerical Modeling of Earth-like Planets\label{sec:models}}

Using maps of the global cloud cover and ice cover 
recorded daily at 0h, 6h, 12h, and 18h UTC\footnote{http://www.ssec.wisc.edu/data/comp/cmoll/},
we simulated Earth-like cloud patterns in addition to
entirely clear atmospheric conditions.

\subsection{Comparison to Unpolarized Satellite Observations of Earth}

We obtained a map of Earth's major land covers from the ISLSCP.\footnote{International Satellite Land Surface Climatology Project;\\ http://daac.gsfc.nasa.gov/CAMPAIGN\_DOCS/FTP\_SITE/INT\_DIS/readmes/veg.html} We converted the ISLSCP land cover
classification to ERBE landforms of Paper 1 as follows.
We treated ISLSCP ``ice'' and ``tundra'' as Paper 1 ``snow,'' ISLSCP ``desert'' as
Paper 1 ``Sahara desert,'' and all other ISLSCP land covers as Paper 1 ``land.''
We interpreted
brighter pixels of the SSEC cloud cover maps as indicative of greater
cloudiness. Over land we allowed four possibilities: clear, partly cloudy,
mostly cloudy, and overcast because those had bi-directional reflectances
parameterized in Paper 1.
We treated ISLSCP ``water'' as Paper 1 ``ocean,'' either the
Dlhopolsky \& Cess (1993)
clear ocean model or the mostly cloudy ocean or overcast as appropriate for
the brightness of the cloud cover map at each particular location. We did not
allow for partly cloudy conditions over oceans simply because to do so would have
required introducing another parameter, the degree of partial cloudiness.

It is worth noting that the specific details of land forms, cloud cover,
etc of our model Earth need not be authentic to be useful, because other worlds
will not be identical to Earth. On the other hand,
the fidelity of our Earth model in reproducing observations of the Earth
(Figure \ref{fig:goesnone}) does contribute to
our confidence in the model's predictive accuracy for potential observations of
Earth-like extra-solar planets.

\subsection{Planets with Earth-like Oceans and Continents: Diurnal Light Curves}

That the short-wavelength light curves of the Earth during one diurnal
cycle in Figure 1 of
Ford, Seager, \& Turner (2001, hereafter Paper 2) match ours
(Figure \ref{fig:fst})
confirms the results of Paper 2 and verifies our numerical methods. 
The diurnal light curve of our model of an Earth with no atmosphere agrees
in detail with that of
Paper 2 (its Figure 1a).\footnote{Figure 1a of Paper 2 corresponds to
an atmosphere-free Earth, not an Earth with a cloud-free atmosphere
as originally stated in Paper 2 (Ford 2006).}
The mean of the two orthogonal linearly polarized reflectances
attributable to Rayleigh scattering from a cloud-free
Earth-like atmosphere at phase $\gamma = 90$\arcdeg\ is $\sim$0.02 in
the units of Figure \ref{fig:fst}, independent of diurnal cycle.
Figure \ref{fig:fst} models the Earth for Sep 22, 2005, viewed from outside the
solar system in the direction, (RA,DEC) = (6$^h$,0\arcdeg). Zero time
corresponds to Julian date 2453636.184. 
We did not simulate the same cloud pattern as Paper 2 did, so Figure 2 of
Paper 2 matches our corresponding diagram only generally:
the mean and variation about the mean of the unpolarized reflectance are
similar in the two simulations. 

In Figure \ref{fig:fst} the {\it difference}\footnote{Hereafter,
an italicized {\it difference} refers to the difference between the
s- and p-components of linearly polarized reflectance.}
between the s- and p-components of linearly polarized reflectance
is anti-correlated with
the total reflectance, because clouds increase the total reflectance while
attenuating the two mechanisms for polarization, glint from the ocean
and Rayleigh scattering of the clear atmosphere.
As the Earth rotates, the {\it difference}
of the s- and p-components is modulated:
the local maxima correspond to clear skies over oceans;
the local minima, to continents or very cloudy regions.
The magnitude of the modulation is approximately one half that
expected from a total obscuration (or not) of a idealized spot with a specular
reflectance given by \ref{eq:speculartimesextinction}.
We attribute the reduced magnitude of the modulation to partial obscuration of 
the specularly reflecting ocean by clouds in our simulated images. 
The region of specular reflection is extended on the globe because the
tilts of the facets of waves spreads the angular distribution of
the specular reflection, which corresponds to a spatial spread on the globe
as observed from space. 

\subsection{Planets with Uniform Surfaces: Phase-dependent Light Curves\label{sec:uniform}}

We simulated planets with uniform ERBE scene types from Paper 1 of
ocean, land, desert, and snow, with or without Earth-like clouds for one
year beginning on Dec 21, 2004 (Figures \ref{fig:ocean} - \ref{fig:pola}).
Representing the data as functions of days since a specific date, instead of
phase, authentically includes the clouds associated with the seasons,
although, for the purposes of this paper, such effects are negligible.
The phase of the Earth advances at $\sim$1\arcdeg\ per day, and
specific phases (full, quarter, new) are indicated on the diagrams.

A ``water-world'' planet with an ocean surface, a clear sky, and no land
shows a prominent
asymmetry about quarter phase in the {\it difference} between the two linear
polarizations (Figure \ref{fig:ocean}). 
Figure \ref{fig:specular} shows
the asymmetric pattern of the {\it difference}
is caused
by the superposition of specular reflection, which is enhanced at crescent
phases, and 
Rayleigh scattering, which is symmetric about quarter phase.
Clouds attenuate the magnitude and the asymmetry of the {\it difference}. 
This magnitude is affected less than the asymmetry, as expected, because
in the models of Paper 1, most of the Rayleigh scattering occurs above
the clouds, whereas the glint is extinguished wherever there are clouds.
In the red and infrared, the specular reflection will persist whereas the
Rayleigh scattering will be attenuated by $\lambda^{-4}$. Hence,
the two diagrams of Figure \ref{fig:specular} can be considered as
approximations to models observed in the red (specular only) or
with no specularly reflecting ocean (Rayleigh only).

For comparison, Figure \ref{fig:specular} includes the 
{\it difference} of the s- and p- components of linearly polarized
reflectance from the analytic formulae of Section \ref{sec:orrery}, including
atmospheric extinction. The similarity of the phase curve of the
{\it difference} from the
analytic formulae (Equation \ref{eq:speculartimesextinction})
to that from the numerical simulations (Section \ref{sec:brdfs}) is remarkable.

\section{Comparison to Earthshine reflected from the Moon\label{sec:earthshine}}

The peak polarization of the Earthshine is $\sim$11\% near
lunar phase angle 80\arcdeg, i.e. a slightly gibbous moon, corresponding to
Earth's phase angle 100\arcdeg\ (Dollfus 1957).
The numerical models described in Section \ref{sec:uniform} predict
that the fractional polarization peaks near
Earth's phase angle $\gamma =$ 100\arcdeg,
in agreement with Dollfus' observations.

The polarization of the Earthshine is less than that of the Earth's light
due to depolarization upon reflection from the Moon (Dollfus 1957).
Adjusting Dollfus' analysis (1957) for a modern value of the
Earth's albedo, A = $0.297 \pm\ 0.005$ (Goode et al. 2001),
we multiple by four times (4.0) the
observed fractional polarization of Earthshine (Dollfus 1957, Fig 25),
to obtain the phase-dependent fractional polarization $p(\gamma)$ of
the Earth's light as seen from space.
The results for $p(\gamma)$ are plotted as filled circles on
Figures \ref{fig:pola} by associating each phase $\gamma$ with
an appropriate date.

\section{Discussion\label{sec:discussion}}

From Equations \ref{eq:lambert}, \ref{eq:specular}, 
\ref{eq:fresnel1}, and \ref{eq:fresnel2},
we predict that an idealized planet with no
atmosphere and an air-water surface illuminated by unpolarized light
will specularly reflect more s-polarized
light than a planet of identical size with an unpolarizing Lambertian
surface of albedo A, for phase angles, $\gamma > \gamma_c$, with 
$\gamma_c =$ 99.2\arcdeg, 110.6\arcdeg, 116.4\arcdeg, 120.3\arcdeg, and
123.2\arcdeg\ for A = 0.2, 0.4, 0.6, 0.8, and 1.0 respectively. 
The corresponding separations from the star are 99\%, 94\%, 90\%, 86\%, and
84\% of maximum separation of a circular orbit. 
Also, the glint's reflectance in s-polarized light exceeds that of
the Lambertian surface for the fraction of a circular orbit,
(180\arcdeg - $\gamma_c$)/180\arcdeg\ = 45\%, 39\%, 35\%, 33\%, and 32\%,
respectively for the five values of albedo.

In principle wind speed may be inferred from the glitter pattern of
specular reflection from sea surface, because as the wind increases, the
wave height does also, which increases the tilts of the facets of the
waves, which broadens the specular reflection in angle compared to that of 
a calm, locally flat surface (Zeisse 1995;
Henderson, Theiler, \& Villeneuve 2003). 
Cox \& Munk (1954)
observed an empirical linear relationship between the local wind speed
and the mean squared slope of waves on Earth's oceans. However
the relationship depends also on a variety of potential factors 
such as the fetch and the duration of the wind, currents in the water, and 
stronger winds (storms) at great distances. That the average
wind speed above the surface of an ocean of an extrasolar planet 
{\it in principle} might be measured is worth noting, perhaps, even if it
seems entirely impractical at the present time. 

The sea-surface glint will increase and decrease
in brightness as ocean and land, respectively, rotate into the localized area
of potential specular reflection.
The small size of the glint area, relative to the radius of the planet,
provides a high-resolution technique to image the boundaries between the
oceans and the continents, if they exist. 
In principle, for some combinations of orbital inclination and planetary
obliquity, this technique could map
the boundaries of continents in two dimensions. Such a map might
provide additional evidence of plate tectonics on the extrasolar planet
in the manner of the western edge of Africa matching the eastern edge
of South America.
Although reconstructing the
continental boundaries from light curves is beyond the scope of this paper,
a one-dimensional example that also includes authentic Earth-like
clouds is presented in Figure \ref{fig:fst}. The oceans and continents are
quite apparent in the {\it difference} between the two linear polarizations.
Temporally and spatially variable clouds will add noise to any
reconstruction of underlying surface features. 
However, we expect reconstructions using the polarized glint could be more
robust than those that
rely solely on non-specularly reflected light for two reasons: 1) the
area of the specular reflection is small compared to the planet's radius,
and 2) the glint is both very bright and strongly polarized, each in a manner
predictable from simple optics, whereas other surface features and
clouds tend not to be strongly polarized. (The Rayleigh-scattered light
is strongly polarized but can be discriminated against by observing in the
red, as we have noted previously).
By examining the {\it difference} of the two orthogonal polarizations
over many diurnal cycles, and over a variety of orbital phases,
one might be able to ascertain the underlying continental boundaries
and the covering fraction of clouds.
In general terms, the technique of
using the localization of the glint for imaging continental boundaries
on extrasolar planets is akin to the method of imaging the
reflected light from a collimated laser beam to overcome turbidity in
the ocean environment (e.g. Moore et al. 2000).

In order to demonstrate the potential utility of the polarized glint from
an Earth-like extrasolar planet, we now estimate its observability. These
estimates are purposefully simplistic in order to permit the reader to
appreciate the advantages, constraints, and limitations of the proposed
technique.
Stam \& Hovenier (2005) similarly analyzed the observability of Rayleigh
scattered light from Jovian planets. 
For both specular reflection and Rayleigh scattering,
the s-component is larger than the p-component, so the two mechanisms
add constructively to the linear polarization of an Earth-like
planet (Figure \ref{fig:goespolarized}).
However, Rayleigh scattering is most significant at short wavelengths whereas
glint is nearly achromatic. Because optical systems tend to perform
better at longer wavelengths, there may be advantages to detecting
glint instead of Rayleigh-scattered light.

A telescope of collecting efficiency $\epsilon$\ and diameter D will
collect photons at a rate ${\dot p}$
from the specular reflection of light 
from a sphere of radius R and reflectivity r at
distance a from a star emitting photons at rate L at
distance d from Earth, where
\begin{equation}
\label{eq:flux}
\dot p \approx {{\epsilon \pi D^2} \over {4}} {{L} \over {4\pi a^2}} {{R^2\ r} \over {4}} {{1} \over {4\pi d^2}}.
\end{equation}
The factors of Equation \ref{eq:flux} are from left to right, the effective area
of the telescope, the star's (photon) irradiance upon the planet,
the specular reflectivity of a sphere (Tousey 1957), and the dilution over
the distance to Earth.
The photon rate from the star directly on the same telescope is
\begin{equation}
\label{eq:fluxstar}
\dot p_* \approx {{\epsilon \pi D^2} \over {4}} {{L} \over {4\pi d^2}},
\end{equation}
and the ratio of the two photon rates is 
\begin{equation}
\label{eq:fluxratio}
\dot p / \dot p_* \approx {{R^2\ r} \over {4}} {{1} \over {4\pi a^2}}.
\end{equation}

For nominal values, D = 10 m, $\epsilon = 0.5$, L = L$_\odot$, a = 1 A.U.,
R = R$_e$, d = 10 pc, the photon rate $\dot p \approx 1.4$ r photons s$^{-1}$.
For simplicity, we assume that the star emits all of its luminosity
at a wavelength near the peak of its spectrum, $\lambda = 0.6\mu$m, but
the ratio in Equation \ref{eq:fluxratio} is independent of that assumption
and equals $3.6 \times 10^{-11}$ times r for the nominal parameters.

In this paragraph, we assume that the planet is observed
such that its glint is much brighter in one polarization than the
other, i.e. at a phase angle $\gamma \ga 90$\arcdeg.
For an air-water interface, the peak difference of the reflectivities
of the two linear polarizations is
$\sim 10$\%, i.e. r $\sim 0.1$ (Figure \ref{fig:orrery3}), which can
be substituted into Equation \ref{eq:flux} to estimate
the excess photon rate in
one polarization compared to the other.
The unpolarized component of the planet's light is negligible with respect to
the star's light scattered in the optics or Earth's atmosphere.
For the nominal parameters above, the angle between the planet and the star,
a/d, is approximately eight times larger than the diffraction width for
the telescope, $\lambda$/D. For sake of discussion, we assume a coronagraph
that can suppress the diffracted star's light at $\sim 8 \lambda$/D
by the factor $\sim 10^{-6}$ with
respect to the on-axis image of a point source. That is $\sim10^{-2}$ times
the local maximum of an Airy pattern at that separation,
and is approximately the geometric mean between demonstrated performance with
ground-based adaptive optics ($\sim 10^{-4}$, Hinkley et al. 2006) and laboratory conditions
in a moderately wide spectral bandpass ($\sim 10^{-8}$, Trauger et al. 2004).
By simultaneously forming two orthogonal, linearly polarized images, or
with an equivalent technique, one may design
the two residual speckle (or diffraction) patterns to be nearly identical
and thus to subtract nearly as well as permitted by Poisson statistics
(however, for technical challenges, see Carson et al. (2006)).
If an imaging polarimeter capable of detecting a fractional polarization of
$\sim 10^{-5}$ could be placed in series with the coronagraph, then
the polarized glint from the an Earth-like planet would be nearly detectable, if
the stellar and instrumental polarizations can be eliminated by design or
by analysis. With respect to the nominal values, if the planet is twice as large,
and twice as near to its star, the planet-to-star flux ratio
would be $0.6 \times 10^{-10}$, or $2^4 = 16$ times larger than nominal,
i.e. $\sim 6$-$\sigma$ greater than the nominal contrast limit of $10^{-11}$.
Because coronagraphs, polarimeters, and telescopes of nearly these
characteristics exist already, and perhaps also do such planets (e.g.
Lovis et al. 2006), some additional advances may
permit detection of oceanic planets around nearby stars
(Schmid et al. 2005, 2006).

In this paragraph we assume the planet is observed with superior instrumentation
that entirely suppresses the star's diffracted light at the location
of the planet.
As derived previously, the photon rate from the planet
$\dot p \approx 1.4$ r photons s$^{-1}$, and r $\approx 0.1$, so $\dot p \approx 0.14 $ photons s$^{-1}$. In an integration time of one hour, a Poisson-limited
signal to noise ratio of $\sim$20 is achieved.
The size of the glint on an Earth-like planet
sets both a spatial and a temporal filter on the
technique of imaging via photometry of the glint: it smooths out surface
features to $\sim 15$\arcdeg\ resolution, corresponding to $\sim 1$ hour
of longitude (Figure \ref{fig:fst}).

One advantage of the polarization of the reflected light is that
the two linear components can be measured separately
and simultaneously. In that case, the difference can be formed,
so that unpolarized light will
cancel out. In the case of a speckle pattern, whether it is variable
on short timescales, such as that of an adaptive optics system, or not,
such as that of an optical system in space, it may be difficult
to determine the {\it sum} of the two polarized components from only the planet
even in cases where the {\it difference} of the two components is detected
well. In that circumstance, we may measure the difference only and will be
unable to determine the fractional polarization of the planet. For this
reason, we emphasize the difference of the two components in this work.

\section{Summary and Future Work}

\begin{figure}
\plottwo{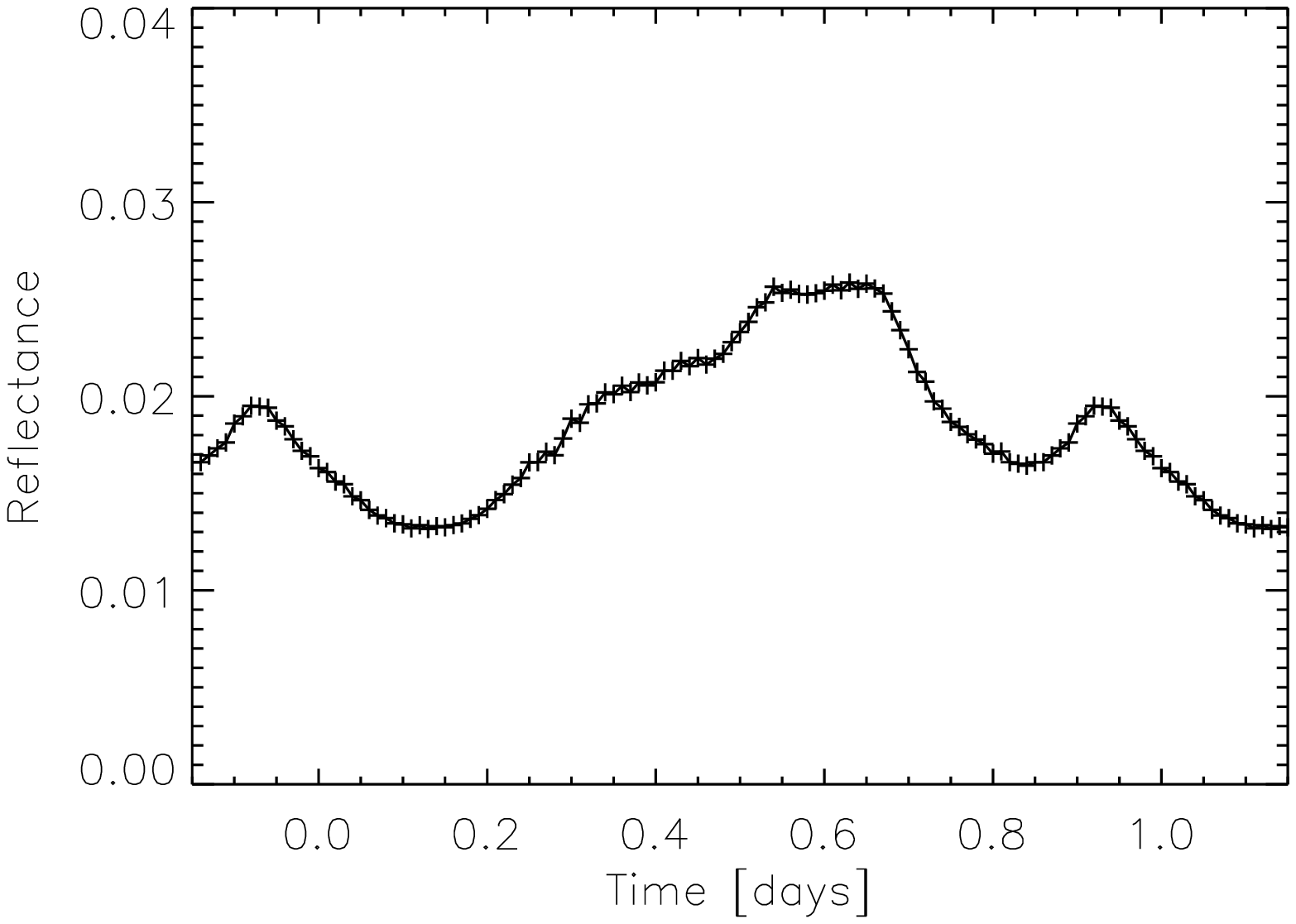}{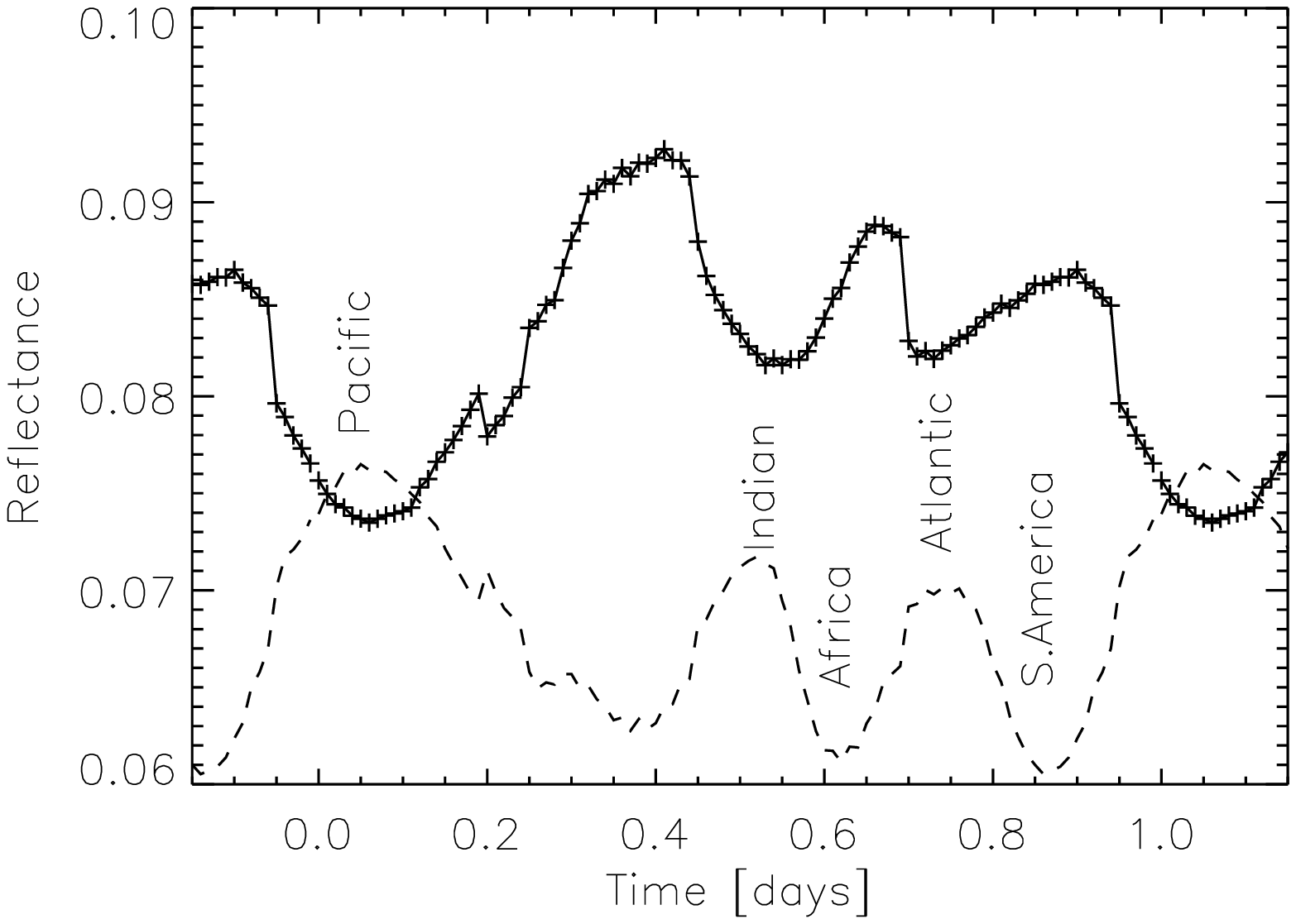}
\caption{Diurnal light curves for the Earth viewed at quadrature.
The reflectance is normalized to 1.0 for a full-phase Lambertian sphere.
Conditions were selected to match those of Figures 1 and 2 of Paper 2. 
Our model of the unpolarized reflectance (left, $+$ symbols) of the
Earth with its atmosphere removed.
The reflectance 
(right, $+$ symbols) for the same conditions except including the atmosphere and
authentic clouds.
The dashed line (right) is twice
the difference between the s- and
p-components of linear polarization.
Major surface features at the location where
the sea-surface glint is or would be, in the case of continents, are
labeled above the dashed line: the local maxima correspond to clear skies
over oceans (Pacific, Indian, and Atlantic, as labeled);
the local minima, to continents (Africa and South America, as
labeled) or cloudy regions (between the Pacific and Indian oceans).
\label{fig:fst}}
\end{figure}

\begin{figure}
\plottwo{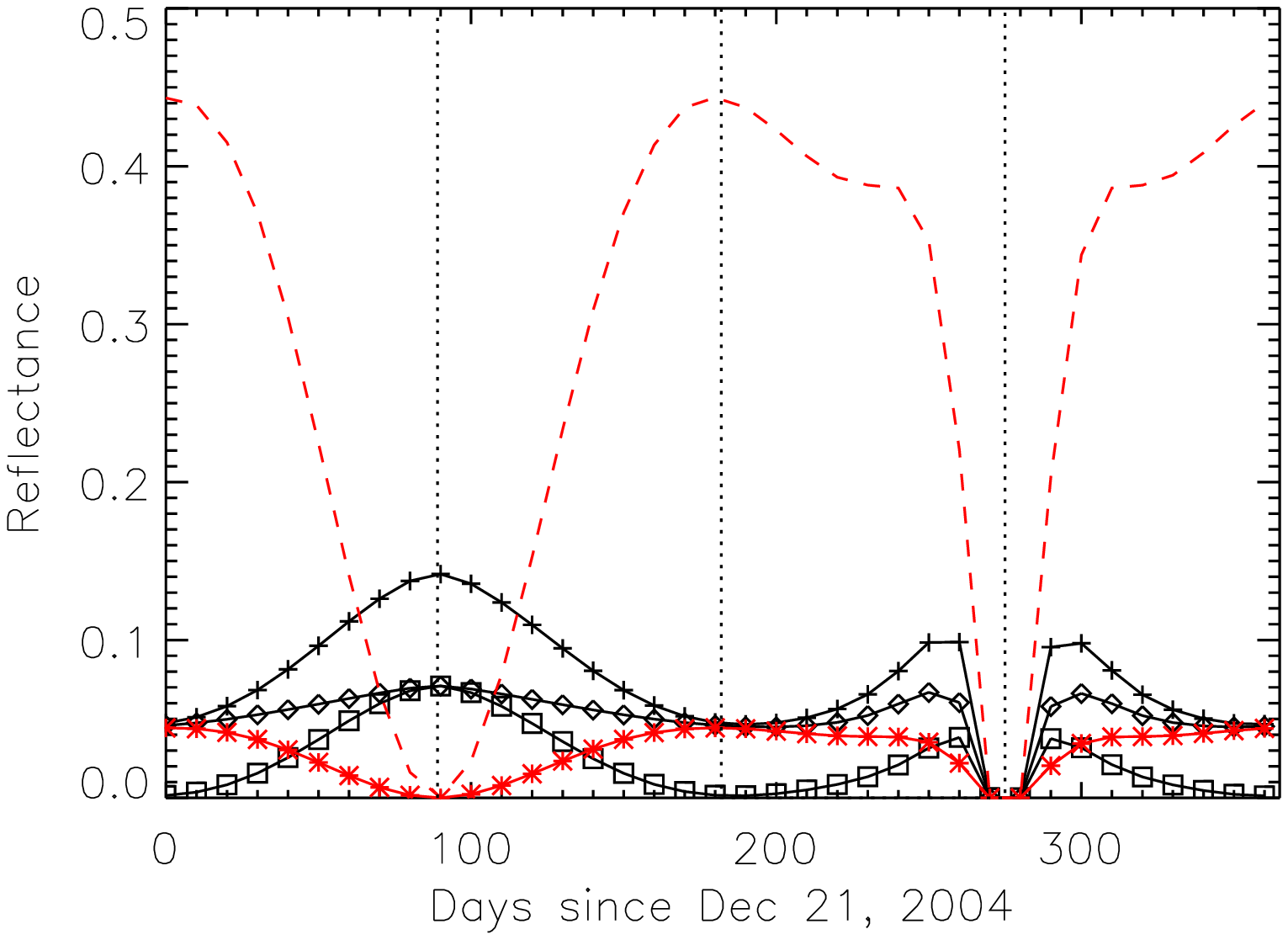}{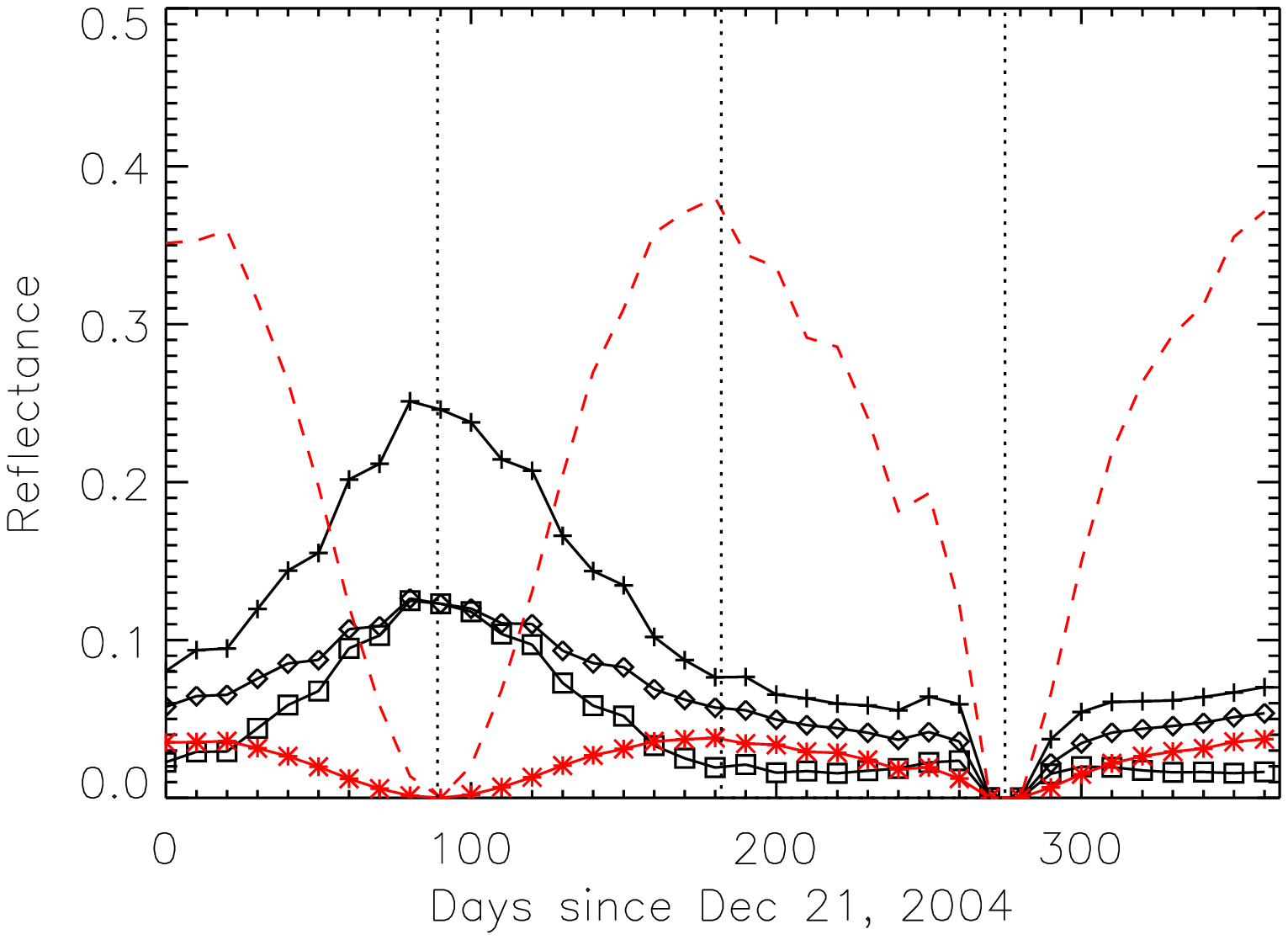}
\caption{Planet with ocean surface. The reflectance is normalized to that
of a Lambertian sphere at its brightest, fully-illuminated phase. The planet's
surface is of a single form, ocean, under an Earth-like atmosphere that is
entirely clear (left) or has clouds with Earth-like
covering fraction and reflectance (right).
The hypothetical observer is located outside the solar system in the ecliptic
plane, in the direction of the vernal equinox.
Dates corresponding to the planet's superior conjunction (i.e. full phase),
quadrature, and
inferior conjunction (i.e. new phase) are indicated by the dotted lines at
89, 182, and 275 days, respectively, after Dec 21, 2004.
The reflectance in the two orthogonal linear polarizations, s and p, are 
indicated by the diamonds and squares, respectively.
Also plotted is the sum ($+$ symbols) of the s and p components and their
difference (asterisk, in red). For clarity ten times the latter is plotted
as a dashed line.
\label{fig:ocean}}
\end{figure}

\begin{figure}
\plottwo{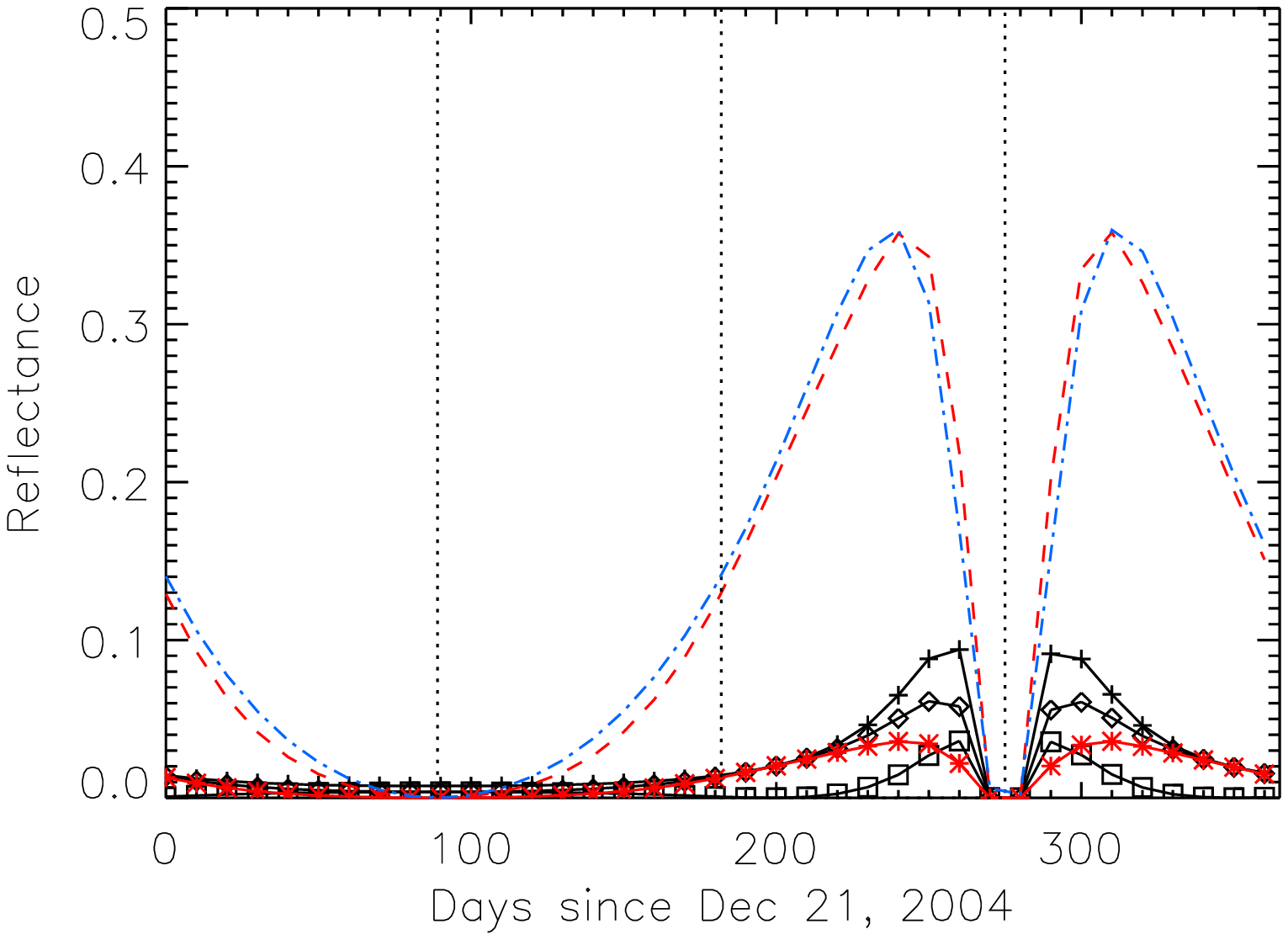}{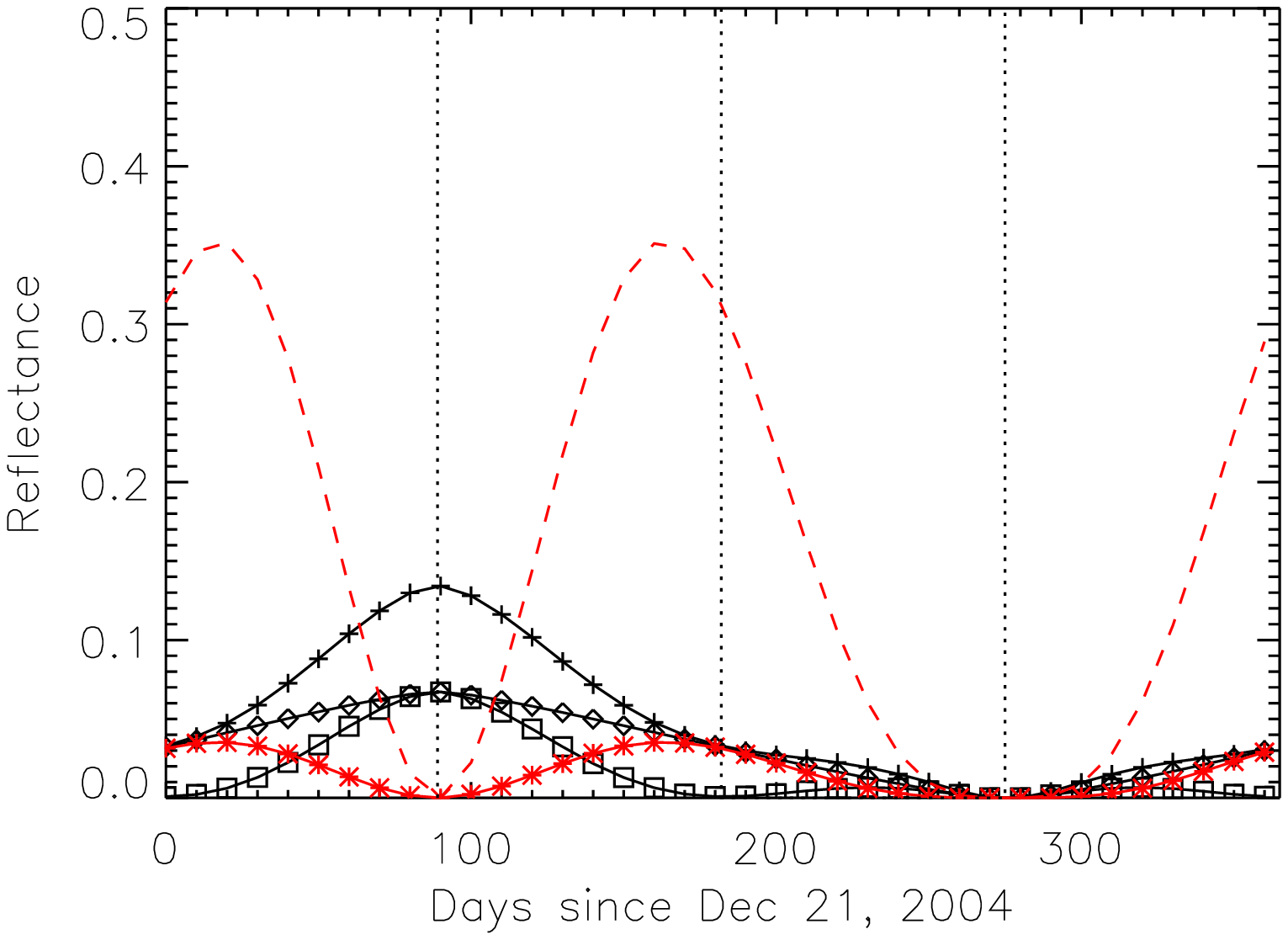}
\caption{Planet with ocean surface and no clouds.
The specular component has been separated
from the others and plotted by itself (left) and removed (right) in order
to aid interpretation of Figure \ref{fig:ocean}.
Legend is the same as Figure \ref{fig:ocean}.
The difference of the s- and p- components of linearly polarized
reflectance from the simple analytic approximations
(Equation \ref{eq:speculartimesextinction} and Figure \ref{fig:orrery3})
is the dash-dot line in the left diagram. The
latter is nearly identical to the equivalent curve generated by numerical
integration of the bi-directional reflectances (dashed line).
\label{fig:specular}}
\end{figure}

\begin{figure}
\plottwo{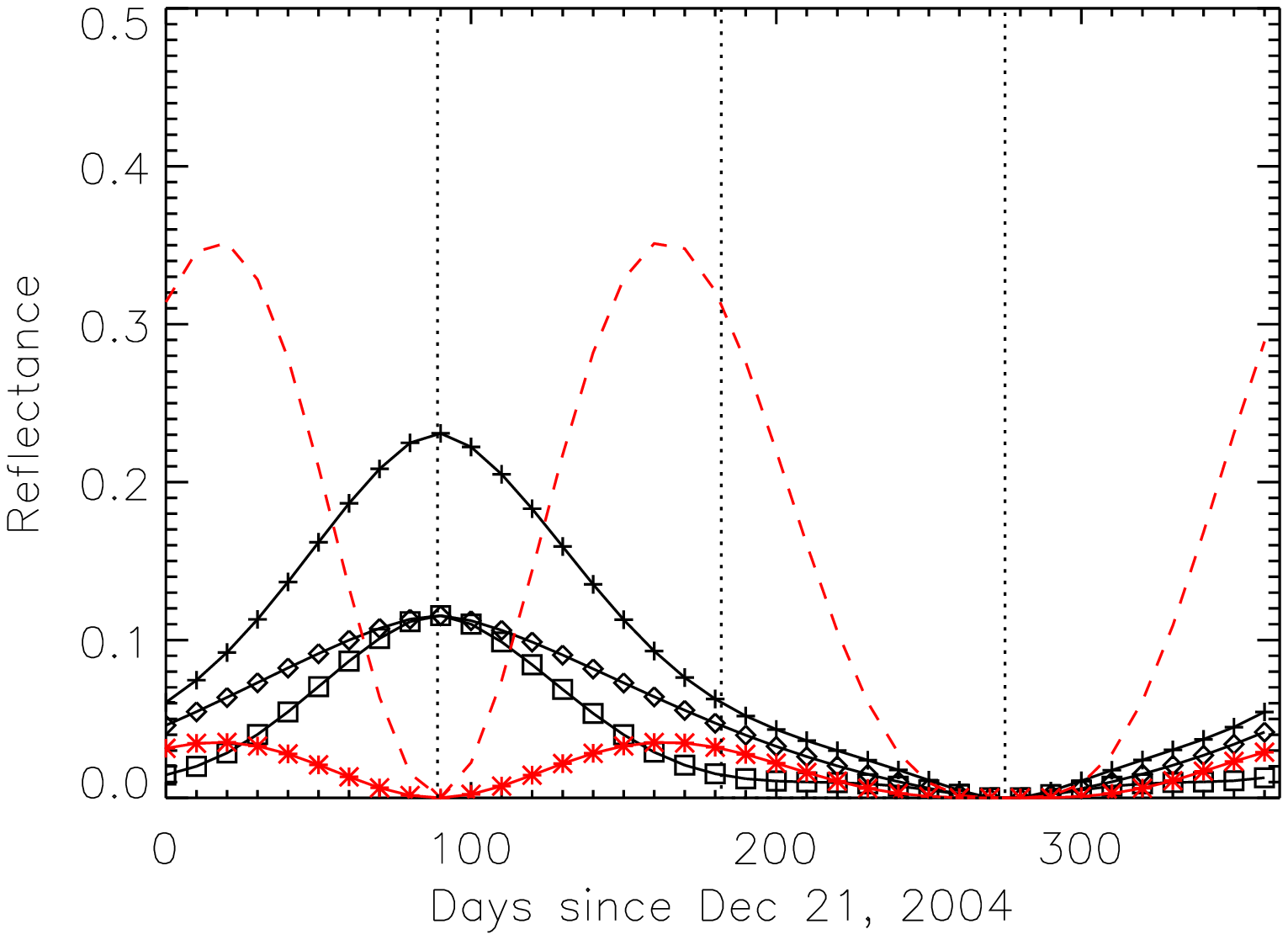}{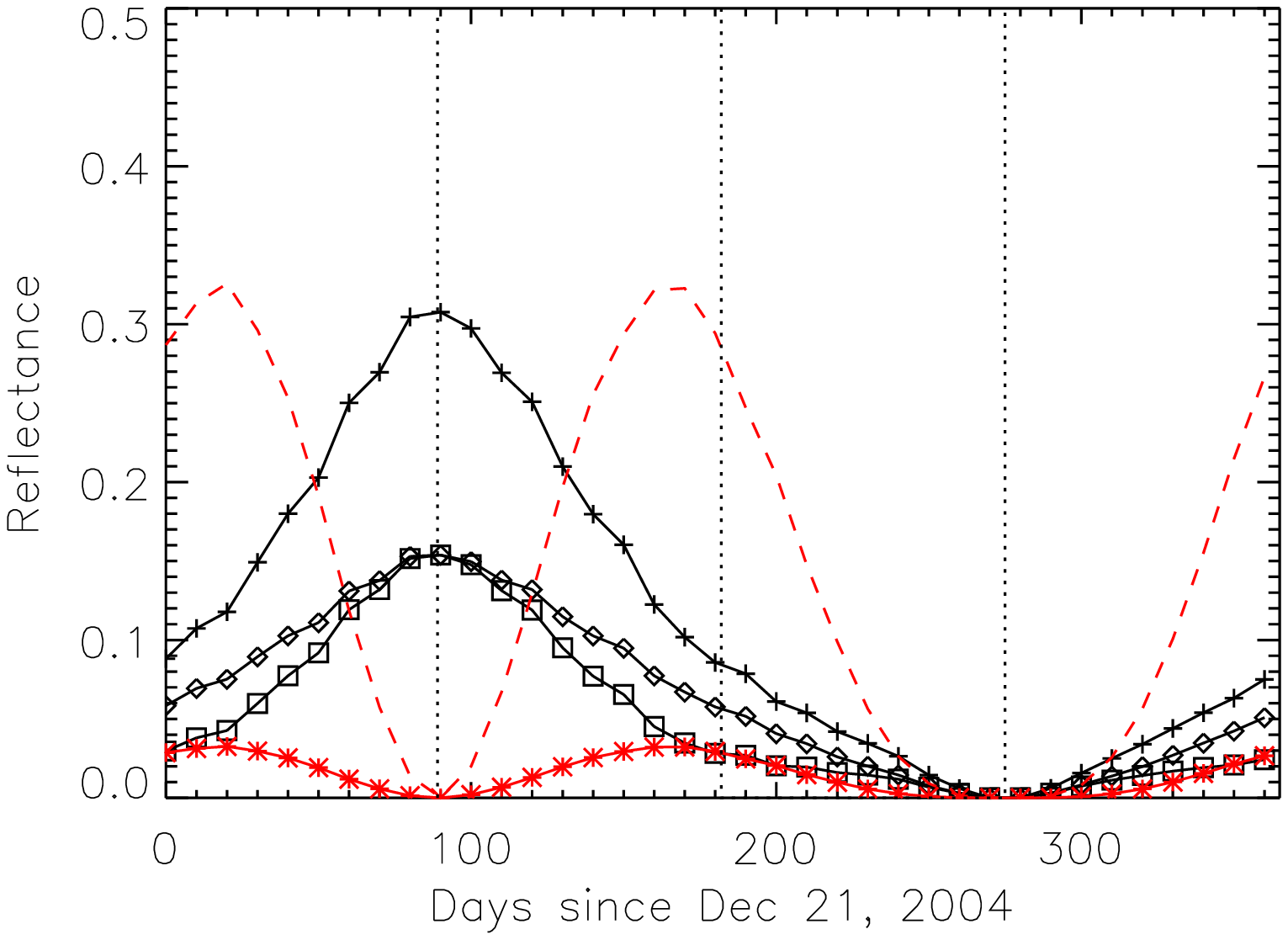}
\caption{Planet with land surface. Legend is the same as Figure \ref{fig:ocean}.
\label{fig:land}}
\end{figure}

\begin{figure}
\plottwo{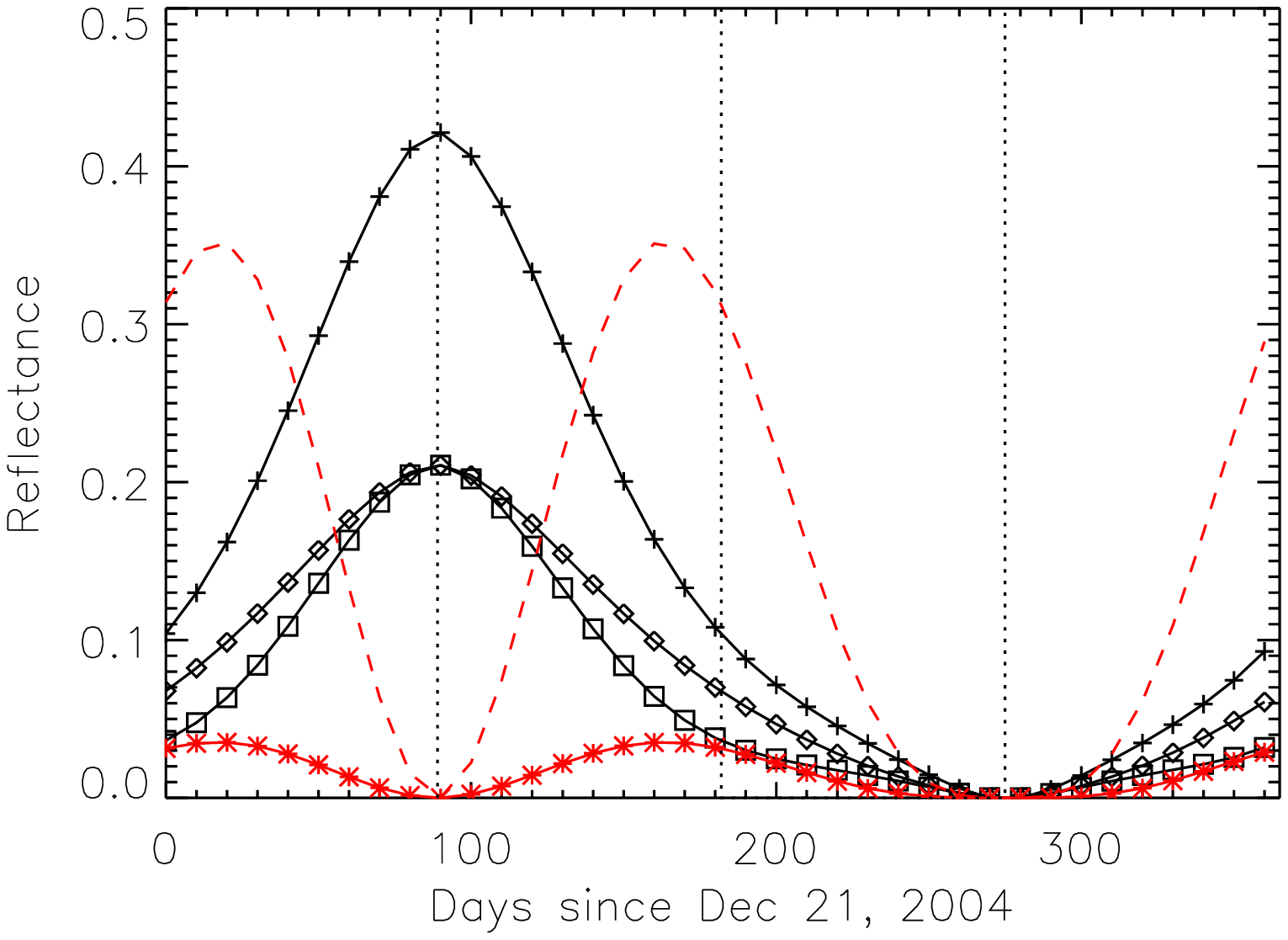}{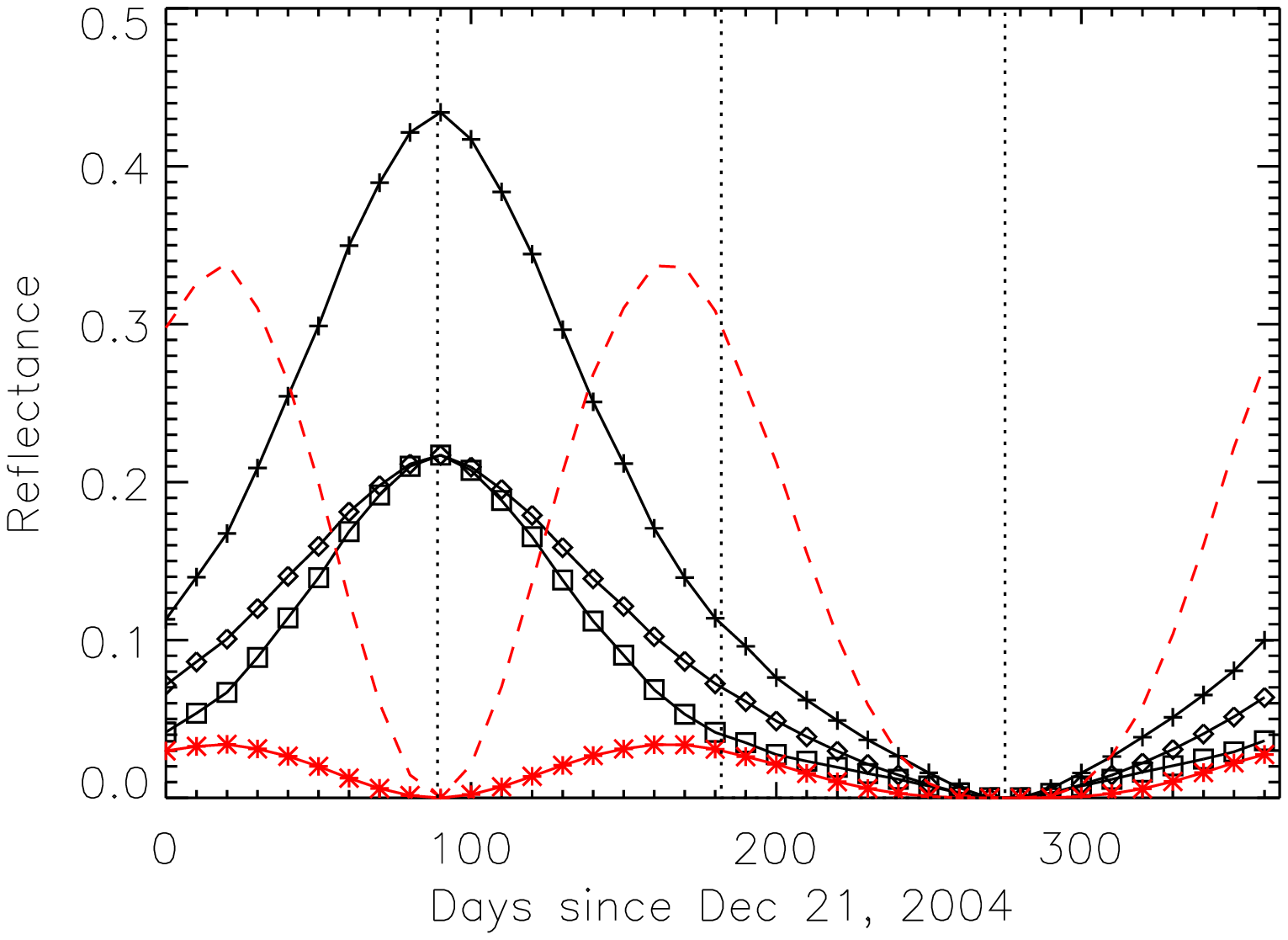}
\caption{Planet with desert surface. Legend is the same as Figure \ref{fig:ocean}.
\label{fig:desert}}
\end{figure}

\begin{figure}
\plottwo{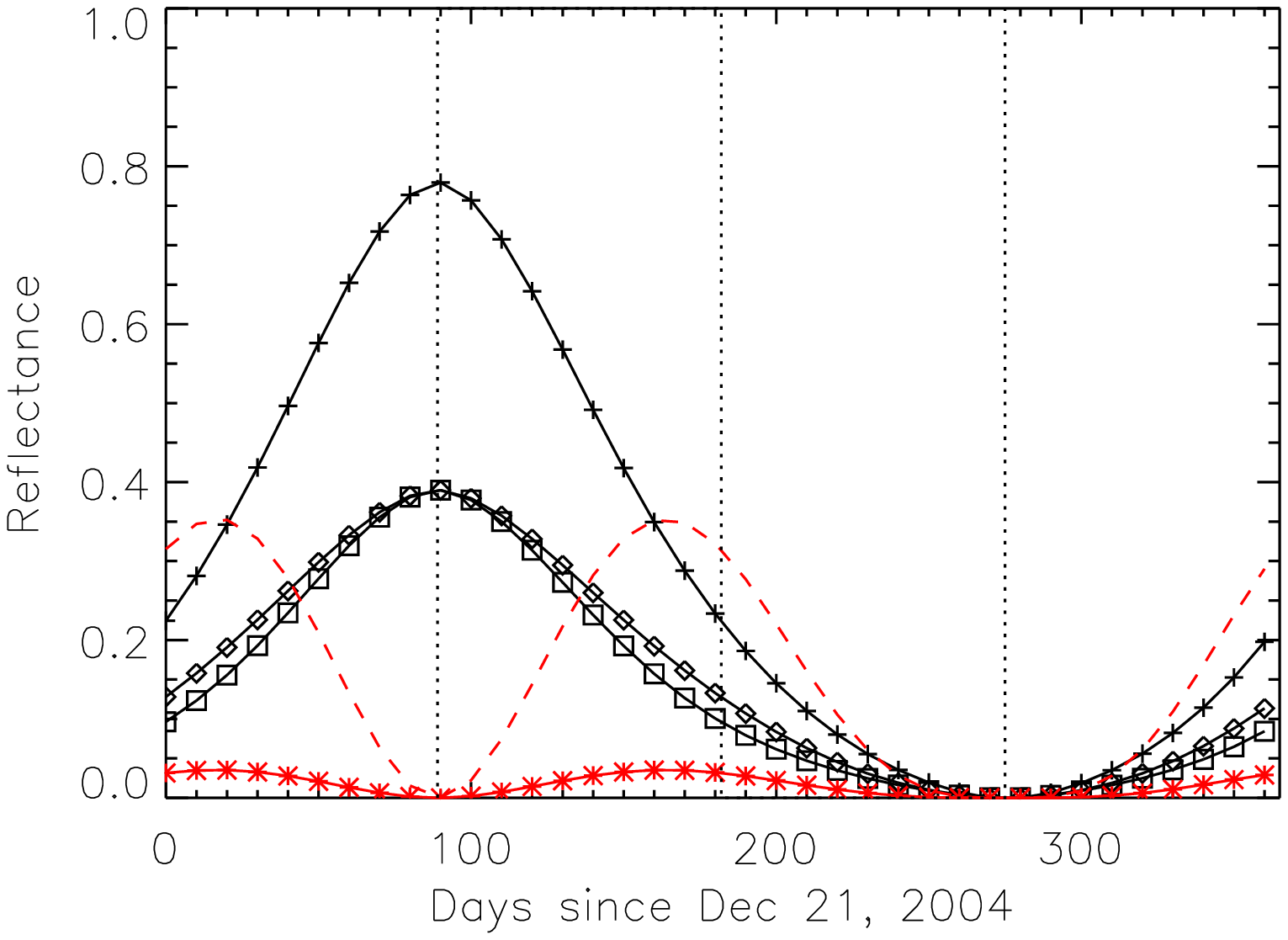}{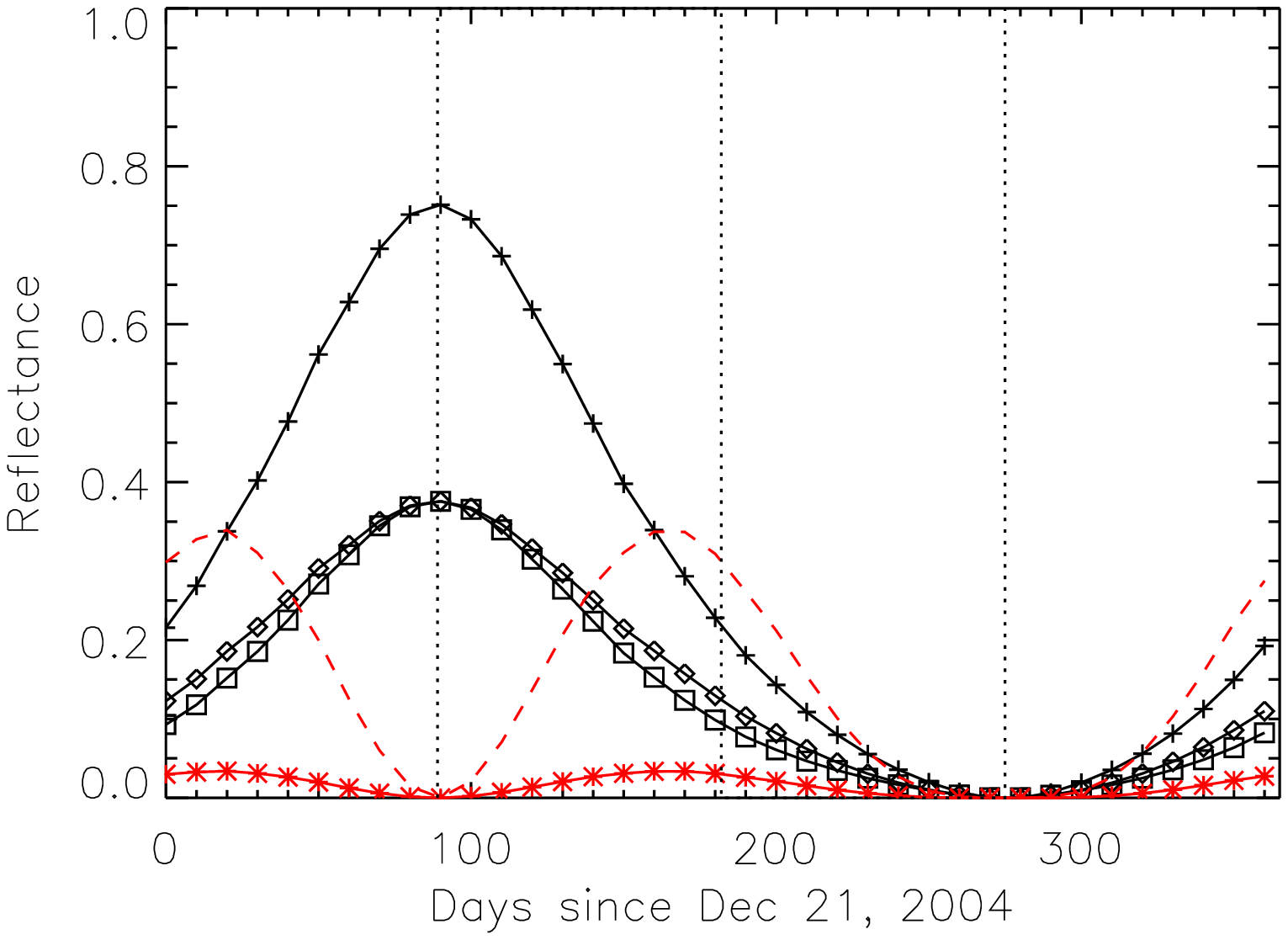}
\caption{Planet with snow surface. Legend is the same as Figure \ref{fig:ocean}.
Due to the large reflectance of snow, the vertical scale is twice that of
the others, Figures \ref{fig:ocean}-\ref{fig:desert}.
\label{fig:snow}}
\end{figure}

\begin{figure}
\plottwo{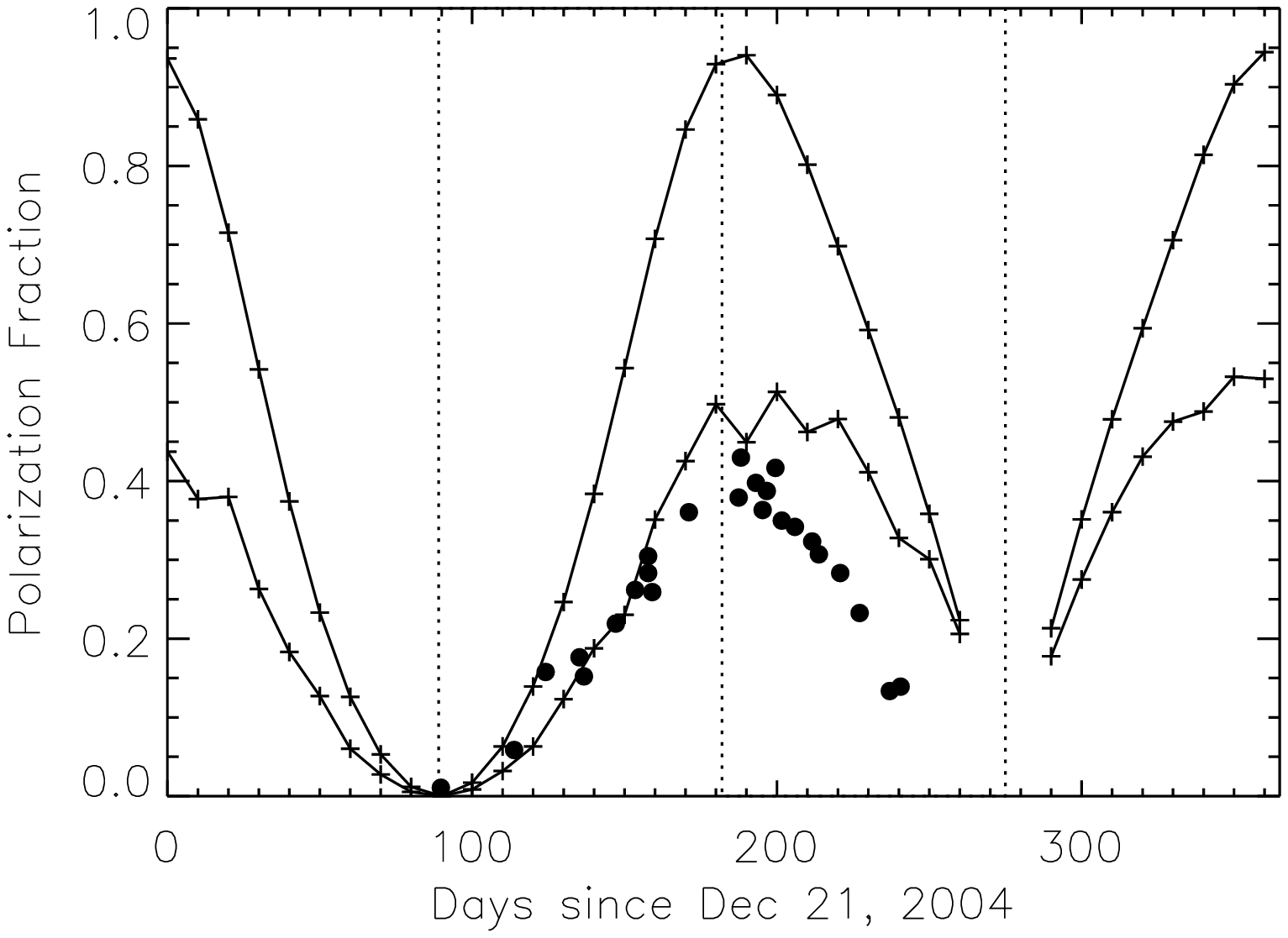}{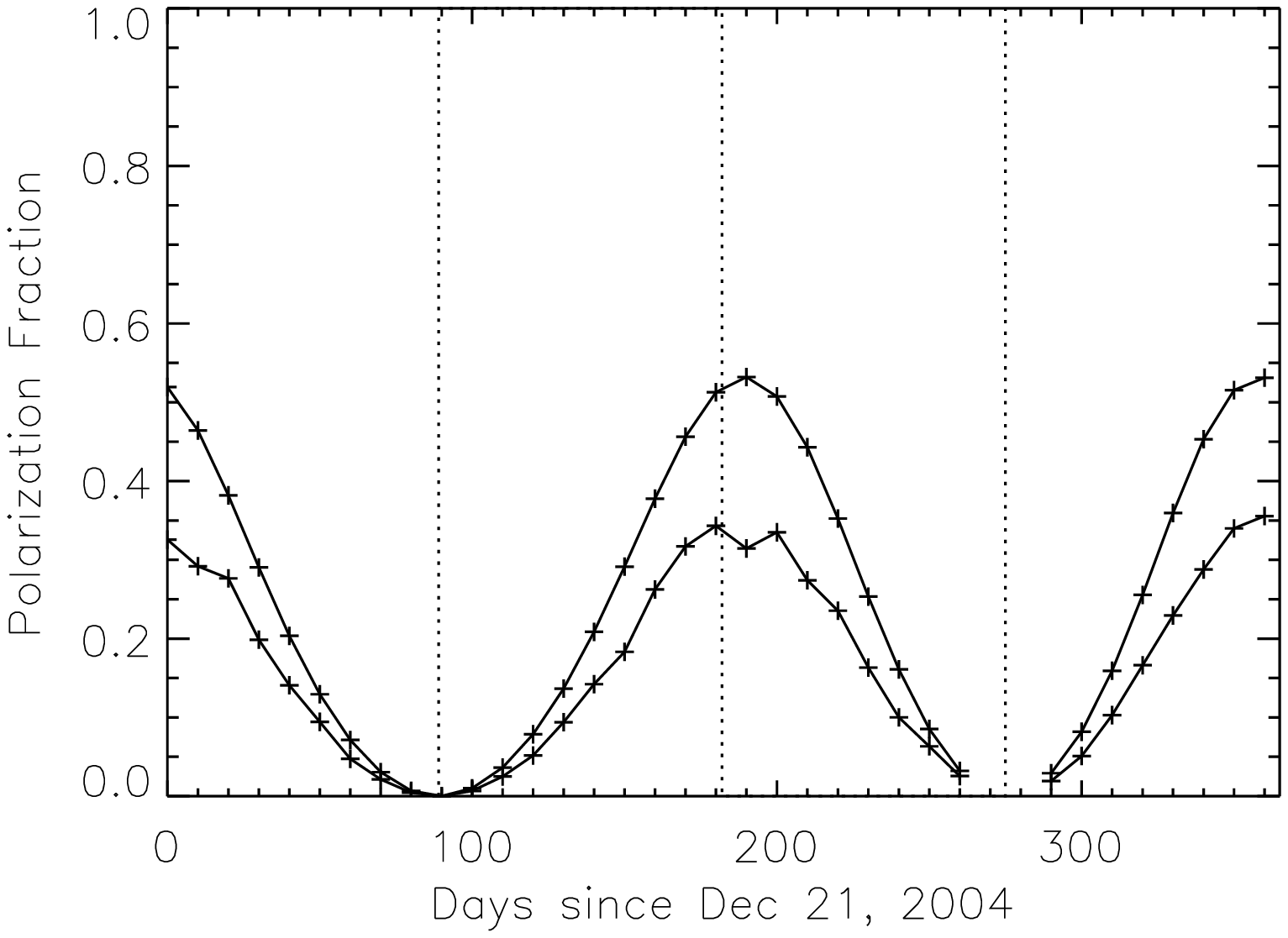}
\plottwo{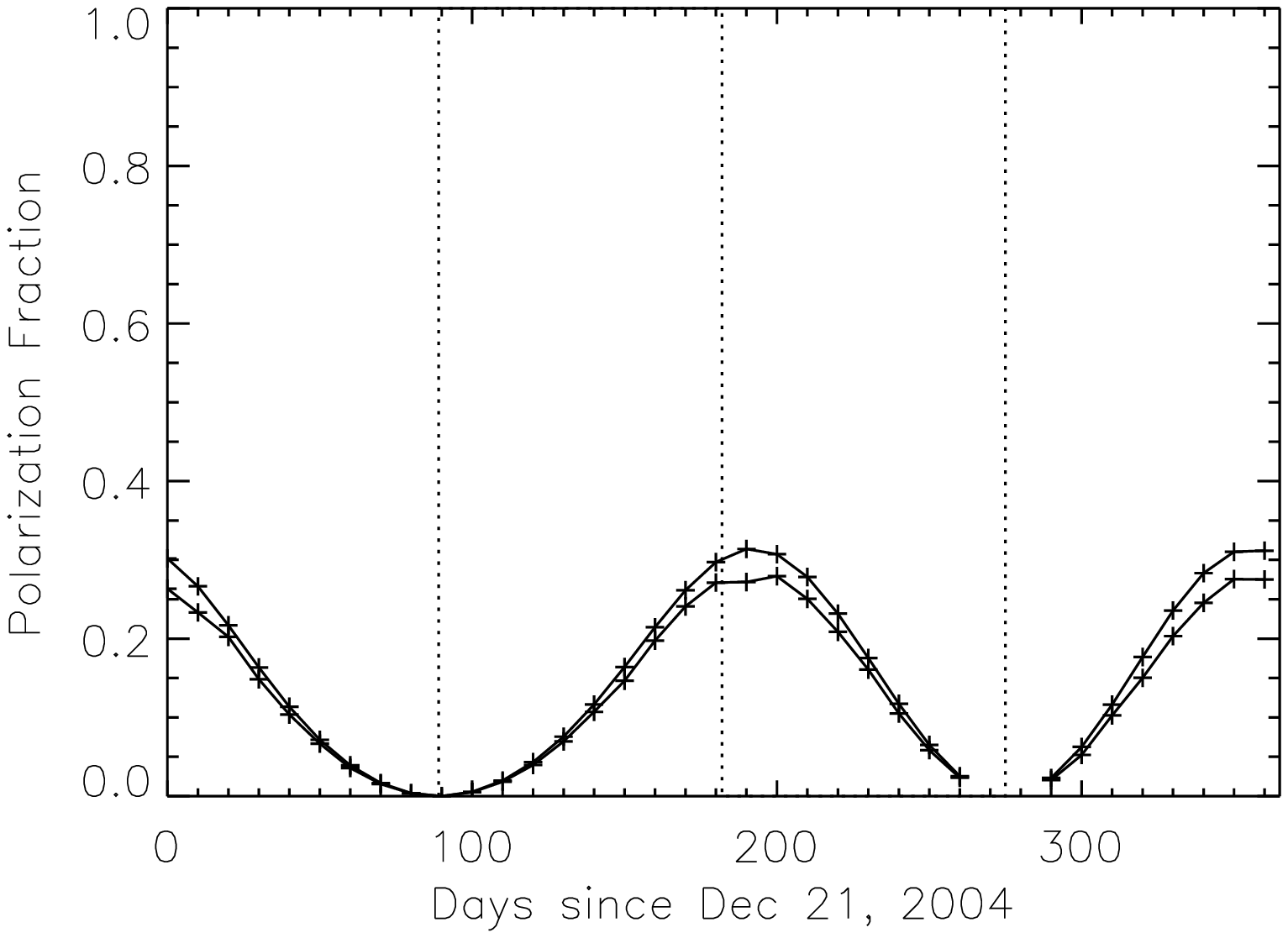}{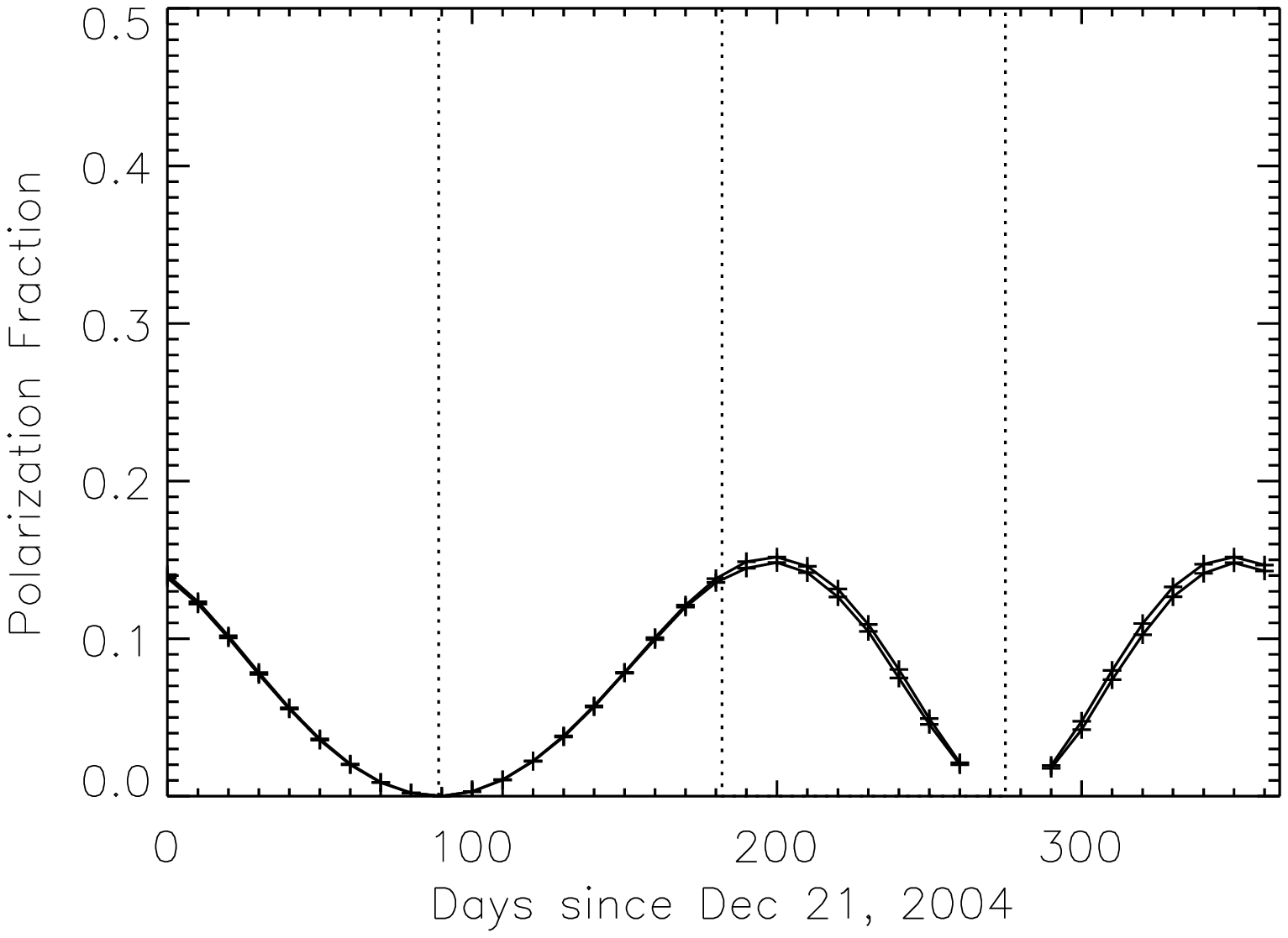}
\plottwo{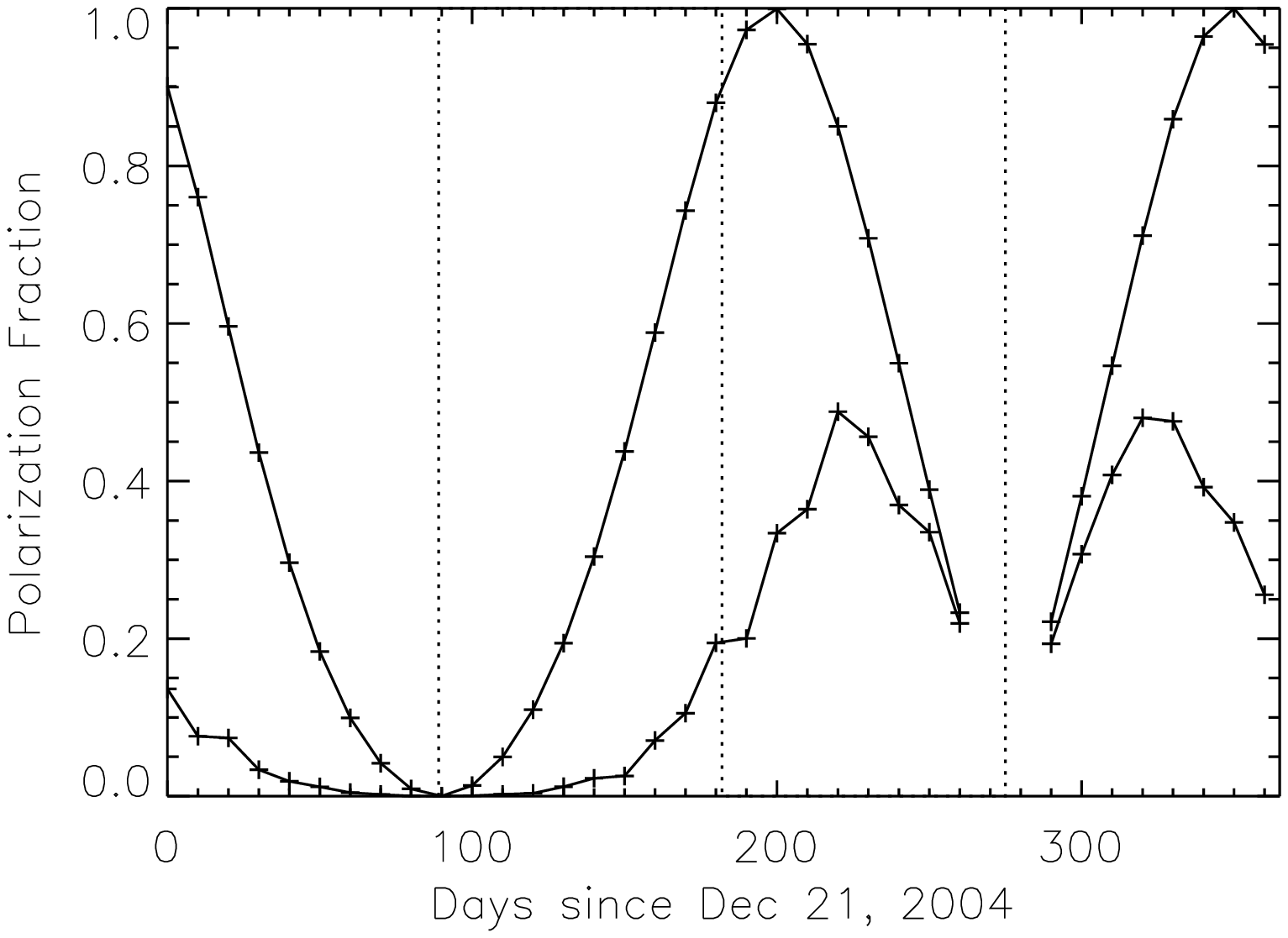}{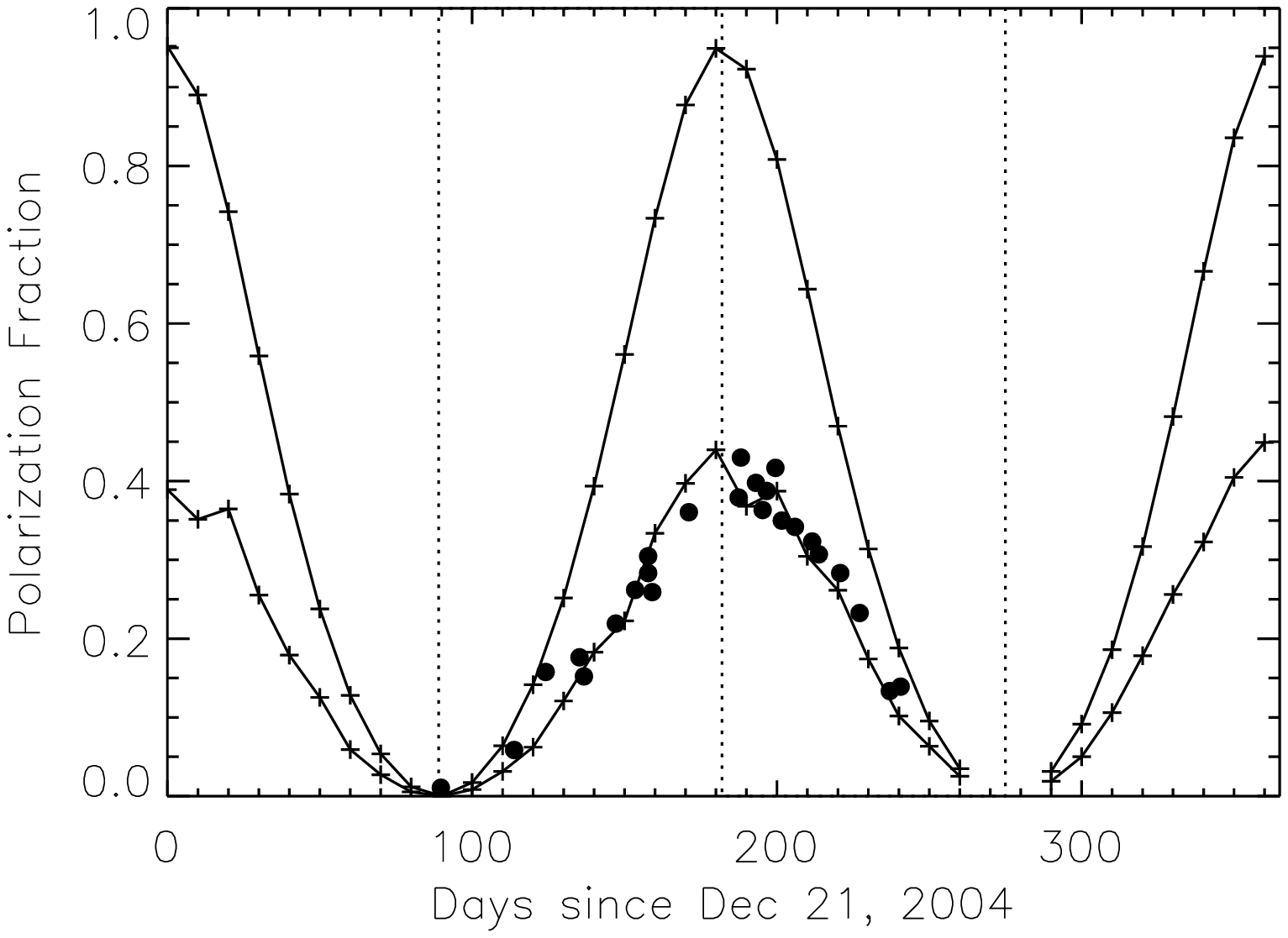}
\caption{Polarization fractions. Planets of various surfaces are simulated
with an Earth-like atmosphere that is
entirely clear (upper curves) or has clouds with Earth-like
covering fraction and reflectance (lower curves).
From left to right and top to bottom, surfaces are ocean, land,
desert, snow, an ocean with only its specular reflection, 
and an ocean with the specular-reflection component eliminated.
Filled circles are data scaled from observations of Earthshine (Dollfus 1957).
\label{fig:pola}}
\end{figure}

This paper predicts the spatially-integrated polarized reflectances of
the Earth and Earth-like planets by simple analytic formulae and by
numerically integrating analytic
approximations for the polarized bi-directional reflectances of
various scene types. Models with and without Earth-like clouds are
compared. Light curves as functions of orbital phase are provided for
an Earth-like planet and for terrestrial planets with a variety of
homogeneous surface types, again with or without Earth-like clouds.
The models predict the {it shape} of the phase function of the
Earthshine's linear polarization observed by Dollfus (1957); the maximum
polarization agrees with Dollfus' observations but is approximately twice as
large as that predicted by Wolstencroft \& Breon (2005).
Simple estimates of the observability of extrasolar Earth-like
planets are provided and indicate that near-future technology
may be able to detect such planets by the difference of planetary
reflectance in two linear polarizations.
Bi-directional reflectances as functions of wavelength would
extend the models described here;
the POLDER satellite can provide such data 
for various scene types (Wolstencroft \& Breon 2005).
We plan to observe the time-variability of the Earth's polarized glint in the
Earthshine to validate the technique of imaging
clouds and continental boundaries described in Section \ref{sec:discussion}
and illustrated in Figure \ref{fig:fst}.

\acknowledgements

Alexander McLane constructed and observed the physical models
described in Section \ref{sec:orrery}.
The author wishes to thank the Distributed Active Archive
Center (Code 902.2) at Goddard Space Flight Center,
Greenbelt, MD, 20771, for reproducing the ISLSCP data on the
web along with IDL code to read it. The original data
products were produced by Dr. Ruth DeFries and Dr. John
Townshend (University of Maryland at College Park,
Department of Geography), with revisions made by Dr.
James Collatz (Code 923, NASA Goddard Space Flight Center).

\clearpage

\end{document}